\documentclass[article]{raa}            

\usepackage{graphicx,times}             
\usepackage{natbib}
\usepackage{amssymb,amsmath}
\usepackage[T1]{fontenc}
\usepackage{CJKutf8}
\bibpunct{(}{)}{;}{a}{}{,}

\def\be{\begin{equation}}
\def\ee{\end{equation}}
\def\CF3{{\sc cosmicflows-3}}
\usepackage{rotating}
\usepackage{multirow}
\usepackage{graphicx}	
\usepackage{amsmath}	
\usepackage{amssymb}	

\usepackage[a4paper=true,pagebackref=true]{hyperref}
\hypersetup{colorlinks = true, linkcolor = green, anchorcolor = red, citecolor = blue, filecolor = red, pagecolor = red, urlcolor = red}

\begin{document}

   \title{Gaussianization of peculiar velocities and bulk flow measurement
}

   \volnopage{Vol.0 (20xx) No.0, 000--000}      
   \setcounter{page}{1}          

   \author{Fei Qin
      \inst{}
   }

   \institute{Korea Astronomy and Space Science Institute, Yuseong-gu, Daedeok-daero 776, Daejeon 34055, Korea; {\it feiqin@kasi.re.kr}\\
\vs\no
   {\small Received~~20xx month day; accepted~~20xx~~month day}}

\abstract{ The line-of-sight peculiar velocities are good indicators of the gravitational fluctuation of the density field. Techniques have been developed to extract cosmological information from the peculiar velocities in order to test the cosmological models. These techniques include measuring cosmic flow, measuring two-point correlation and power spectrum of the peculiar velocity fields, reconstructing the density field using peculiar velocities. However, some measurements from these techniques are biased due to the non-Gaussianity of the estimated peculiar velocities. Therefore, we use the 2MTF survey to explore a power transform that can Gaussianize the estimated peculiar velocities.  We find a tight linear relation between the transformation parameters and the measurement errors of  log-distance ratio. To show an example for the implement of the Gaussianized peculiar velocities in cosmology, we develop a bulk flow estimator and estimate bulk flow from the Gaussianized peculiar velocities. We use 2MTF mocks to test the algorithm, we find the algorithm yields unbiased measurements. We also find this technique gives smaller measurement errors compared to other techniques. 
Under the Galactic coordinates, at the depth of $30$ $h^{-1}$ Mpc,
we measure a bulk flow of $332\pm27$ km s$^{-1}$ in the direction $(l,b)=(293\pm 5^{\circ},  13\pm 4^{\circ})$. The measurement is consistent with the $\Lambda$CDM prediction.
}

   \authorrunning{F. Qin }            
   \titlerunning{Bulk flow from the Gaussianized 2MTF PVs }  

   \maketitle

%
%
\section{Introduction}

Driving by the expansion of the Universe, galaxies move further apart from us. This motion is called recessional velocity and described by the Hubble's Law which is a linear relation between the redshift and distance of galaxies. On small scales, the mass density field of the Univers is not ideally homogenous and isotropic, which results from gravitational fluctuation. On top of the Hubble recessional velocities, galaxies will have peculiar motions which arise from these gravitational perturbations of the mass density field. The line-of-sight peculiar velocities of galaxies enable us to test the cosmological models through three main techniques. 

One technique directly measures the cosmic flow field using the peculiar velocities, then comparing to the cosmological models' prediction to test whether the models accurately describe the motion of galaxies. Some examples of the previous work related to this method are
\citealt{Kaiser1988,Lister1989,Jaffe1995,Nusser1995,Parnovsky2001,Nusser2011,2012MNRAS.420..447T,MA2012,Ma2013,Ma2014,Hong2014,Scrimgeour2016,Qin2018,Qin2019a,Boruah2020}. The measurements agree with the  
$\Lambda$ cold dark matter ($\Lambda$CDM) model prediction.

The second technique measures the two-point correlation and/or power spectrum of the peculiar velocity field and fits the cosmological parameters, then comparing to the cosmological models' prediction. Some examples of the previous work related to this method are \citealt{Gorski1989,Kolatt1997,Zaroubi1997,Juszkiewicz2000,Silberman2001,Feldman2003,Gordon2007,Johnson2014,Howlett2017,Huterer2017,Dupuy2019,Howlett2019,Qin2019b}. 

The third technique is the reconstruction of the density/velocity field of the local Universe using peculiar velocities. Some examples of the previous work related to this method are \citealt{Nusser1994},\citealt{Erdogdu2006a},\citealt{Lavaux2010},\citealt{Springob2014}
,\citealt{Carrick2015},\citealt{Pomar2017},\citealt{Springob2016}. 

In most of the past literature, the measurement errors of peculiar velocities are assumed to be Gaussian which is not true for the usual peculiar velocity estimator. This can bias the measurements, and many researches have been done to deal with the non-Gaussianity of the estimated peculiar velocities.
For example, in terms of cosmic flow measurements, to avoid the non-Gaussianity of the estimated peculiar velocities, \citealt{Nusser1995,Nusser2011,Qin2018,Qin2019a} use the so-called $\eta$MLE to measure the cosmic flow in the logarithmic distance ratio-space. \citealt{Watkins2015} developed a peculiar velocity estimator which has Gaussian errors but biased in some circumstances. In terms of power spectrum measurements,  \citealt{Qin2019b} use a power transformation to offset the non-Gaussianity of the momentum power  spectrum \citep{Howlett2019} to fit the growth rate of the large scale structure.

In this paper,  we will explore a technique that Gaussianizes the estimated line-of-sight peculiar velocities. In addition,
we take the bulk flow measurement as an example to show the implementation of the Gaussianized peculiar velocities in terms of testing cosmology. The survey data used in this paper is a full-sky Tully-Fisher survey 2MTF \citep{Hong2019}.

The paper is structured as follows: 
in Section \ref{sec:data} we introduce the 2MTF data and mocks. The mocks are used to test the algorithm. 
In section \ref{sec:PVest} we introduce the peculiar velocity estimators and discuss the Gaussianity of the estimated peculiar velocities.
In section \ref{sec:box} we introduce the algorithm used to Gaussianize the peculiar velocities.
In section \ref{sec:BKest}  we introduce the bulk flow estimator, which estimates bulk flows from Gaussianized peculiar velocities and test the estimator using mock surveys.
In section \ref{sec:dis} we present the bulk flow measured from 2MTF. A conclusion in presented in Section \ref{conc}.

This paper assumes spatially flat cosmology. The cosmological parameters used in this paper are from the \cite{Planck2016A}: $\Omega_m=0.307$, $\Omega_{\Lambda}=0.693$, $\sigma_8=0.823$ and $H_{0} = 100$ $h$ km s$^{-1}$ Mpc$^{-1}$, $h=0.678$. These parameters are applied to the calculation of the comoving distances and the  bulk flow predicted in $\Lambda$CDM .

\section{DATA AND MOCKS} \label{sec:data}

\begin{figure*} 
 \includegraphics[width=75mm]{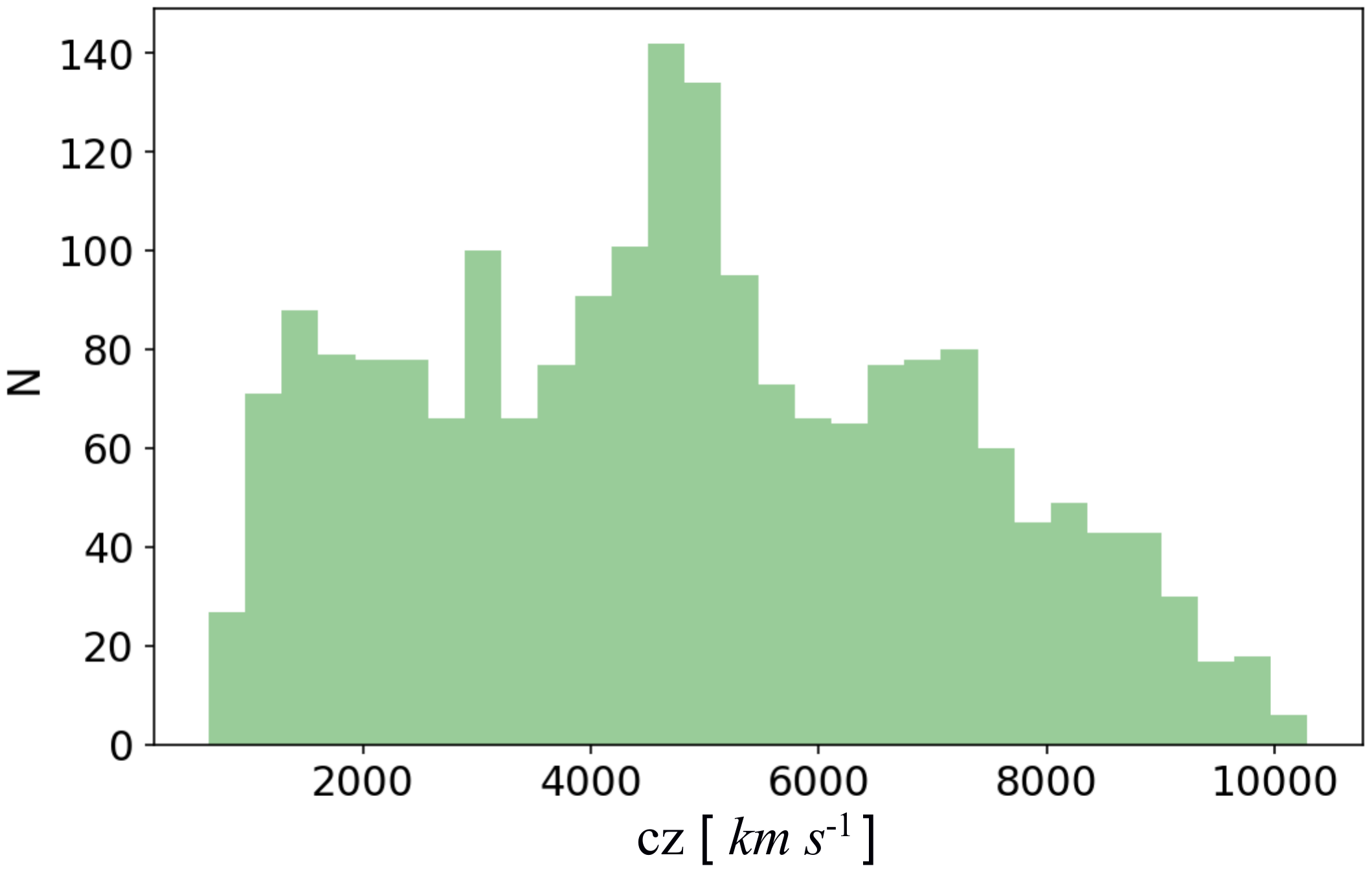}
  \includegraphics[width=100mm]{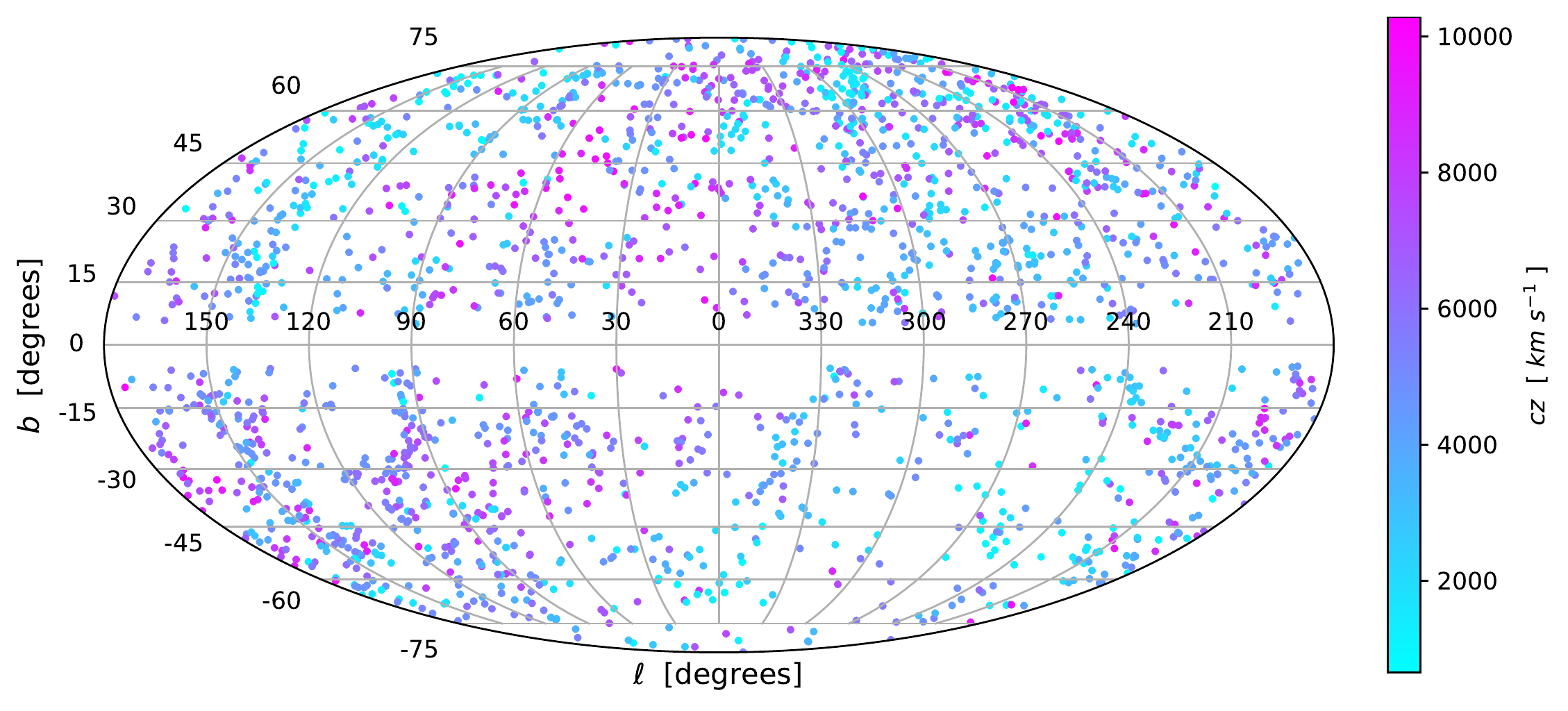}
 \caption{The survey geometry of 2MTF. The left-hand-side panel shows the distribution of redshift of 2MTF galaxies. The right-hand-side panel shows the sky coverage of the 2MTF galaxies, the color of the points indicates the galaxy redshift, according to the colour bar.}
 \label{lb}
\end{figure*}

2MTF \citep{Hong2019} is a full sky Tully-Fisher
survey derived from the Two Micron All-Sky Survey (2MASS, \citealt{Masters2008,Huchra2012,Hong2014}). 
The redshift of 2MTF galaxies reaches a maximum of $1.2 \times10^{4}$ km~s$^{-1}$ and no smaller than 600 km s$^{-1}$. 2MTF is a full-sky survey, but excluding the  Galactic plane region where Galactic latitude $|b|<5^{\circ}$.  Fig.\ref{lb} shows the survey geometry (redshift distribution and sky coverage) of the 2MTF galaxies.

\begin{figure} 
\centering
 \includegraphics[width=90mm]{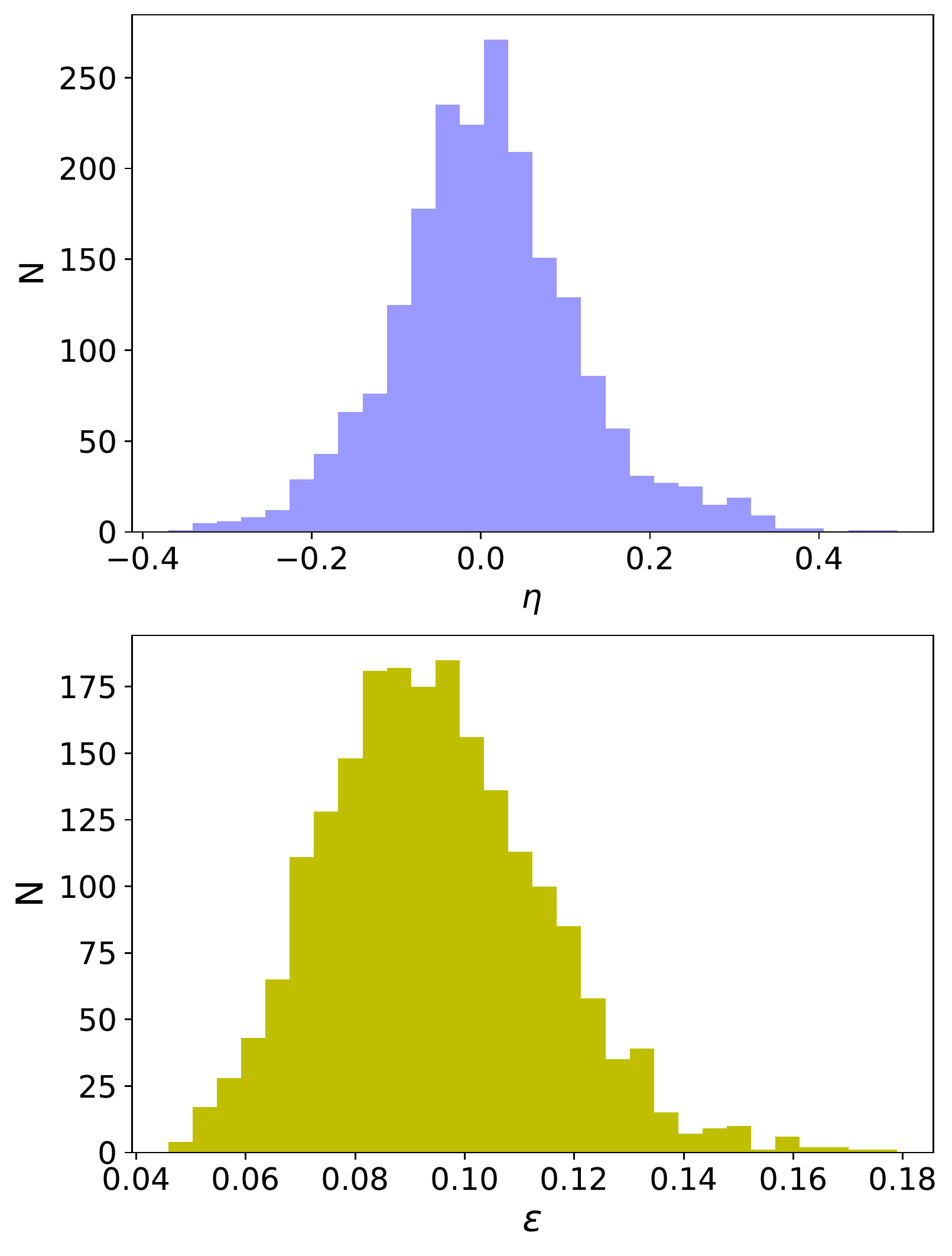}
 \caption{The top panel shows the distribution of log-distance ratio $\eta$ of 2MTF. The bottom panel shows the distribution of measurement errors of log-distance ratio $\epsilon$.}
 \label{logds}
\end{figure}

The logarithmic distance (log-distance) ratio for a galaxy is defined as
\be\label{logd}
\eta \equiv \log_{10}\frac{d_z}{d_h}~,
\ee
where $d_z$ is the apparent comoving distance of a galaxy inferred from its observed redshift $z$, and $d_h$ is the true comoving distance of the galaxy. In the 2MTF survey, $\eta$ is estimated from the Tully-Fisher relation \citep{Masters2008,Hong2014}. The distribution of the log-distance ratio of the 2MTF galaxies is shown in the top panel of Fig.\ref{logds}, and the measurement error of log-distance ratio, $\epsilon$, is shown in the bottom panel of Fig.\ref{logds}.

We use 16 mock 2MTF catalogues to test the algorithm used  in this paper. These mocks are not used in the real data analyse and the comparison of $\Lambda$CDM.  The mock sampling algorithm is clearly presented in \citealt{Howlett2017}. 2MTF has a well defined selection function \citep{Hong2014}, enabling us to  generate high quality mocks which can accurately realize the survey geometry and selection function of 2MTF \citep{Howlett2017,Qin2018,Qin2019a,Qin2019b}. In this paper the 16 mocks are generated from two different simulations: the GiggleZ   \citep{Poole2015} and the SURFS \citep{Elahi2018} simulations. The cosmological parameters used in the GiggleZ simulation are $\Omega_m=0.273$ and $h=0.705$, while for the SURFS simulation, $\Omega_m=0.3121$ and $h=0.6751$. 
Two different simulations are used to produce the mocks, enabling us to ensure that the algorithm presented in this paper gives consistent answers for different cosmologies \citep{Qin2018}.

\section{Peculiar Velocity estimators} \label{sec:PVest}

If neglect the relativistic motions and gravitational lensing effects, the line-of-sight peculiar velocity of a galaxy, $V$,  can be estimated from its log-distance ratio $\eta$ and observed redshift $z$ through \citep{Colless2001,Hui2006,Davis2014,Scrimgeour2016,Qin2018,Qin2019a}
\be \label{travp}
V=c\left(\frac{z-z_h}{1+z_h}\right)~,
\ee
where $c$ is the speed of light. The Hubble recessional  redshift $z_h$ is numerically calculated from the true comoving distance, $d_h$ of the galaxy using
\be\label{Dz}
d_{h}(z_{h})=\frac{c}{H_0}\int_0^{z_{h}}\frac{dz'}{E(z')}~,
\ee
where
\be\label{Ez}
E(z)=\frac{H(z)}{H_0}=\sqrt{\Omega_m(1+z)^3+\Omega_{\Lambda}}~,
\ee
where $H_{0}$, $\Omega_{m}$ and $\Omega_{\Lambda}$ are the Hubble constant, matter and dark energy densities of the present day universe, respectively. The true comoving distance
\be \label{etadist}
d_{h} = d_{z}10^{-\eta} ,
\ee
where the apparent comoving distance $d_z$ is calculated from the observed redshift $z$ directly  through a similar expression of Eq.~\ref{Dz}. 

For a galaxy, Eq.~\ref{travp} converts $z$ and $\eta$ to $V$ non-linearly. Therefore, the measurement error of $V$, which is propagated from the measurement error of $\eta$, is not Gaussian, even if we assume   $\eta$ has Gaussian error. To see this clearly, using equations ~\ref{travp}, \ref{Dz} and \ref{Ez}, one can calculate the probability distribution function (PDF) of a estimated line-of-sight peculiar velocity, given by \citep{Scrimgeour2016}
\be \label{PDFv}
P(V)=P(\eta)\frac{d\eta}{dV}=P(\eta)\times \frac{(1+z_h)^2}{d_hH_0E(z_h)(1+z)\ln(10)}~,
\ee
where $P(\eta)$ denotes the PDF of $\eta$. Usually (but not necessarily), $P(\eta)$ is assumed to be a Gaussian function. However, due to the non-linear term behind $P(\eta)$, the resultant $P(V)$ is not Gaussian. 
Therefore, the peculiar velocity estimated from Eq.~\ref{travp} for a galaxy does not have Gaussian error (see Section \ref{sec:box1} and Fig.\ref{PVss} for more discussions). 

Due to the non-Gaussianity of the peculiar velocities, the  cosmological parameters and cosmic flows estimated from the peculiar velocity fields are biased to some extent.  \citealt{Qin2019a} finds that this non-Gaussianity will bias the momentum power spectrum measurements and then  bias the estimation of the growth rate of the large-scale-structure. Using simulations, \citealt{Qin2018} shows that this non-Gaussianity will bias the cosmic flow measurements comparing to the true values.

To preserve the Gaussianity of the estimated peculiar velocities, \cite{Watkins2015} developed the following estimator to calculate peculiar velocity for a galaxy 
\be\label{watvp}
V=\frac{cz_{\mathrm{mod}}}{1+z_{\mathrm{mod}}}\ln\frac{cz_{\mathrm{mod}}}{H_0d_l}~,~~(V_t\ll cz)~,
\ee
where $d_l$ denotes the luminosity distance, and $z_{\mathrm{mod}}$ is given by
\be  \label{zmod}
z_{\mathrm{mod}}=z\left[1+\frac{1}{2}(1-q_0)z-\frac{1}{6}(1-q_0-3q_0^2+1)z^2\right].
\ee
where $q_0=0.5(\Omega_m-2\Omega_{\Lambda})$ is the so called acceleration parameter. At low redshift, 
$\ln\frac{cz_{\mathrm{mod}}}{H_0d_l} \approx \ln(10)\eta$, then the PDF for a peculiar velocity estimated from Eq.\ref{watvp} is given by:
\be \label{PDFv2}
P(V)=P(\eta)\frac{d\eta}{dV}=P(\eta)\times \frac{1+z_{\mathrm{mod}}}{cz_{\mathrm{mod}}\ln(10)}~.
\ee
Assuming $P(\eta)$ is Gaussian, for a given galaxy, $z_{\mathrm{mod}}$ is a certain number, $P(V)$ is linear related to $P(\eta)$ and therefore is Gaussian. Therefore, the peculiar velocity estimated from Eq.\ref{watvp} for a galaxy does have Gaussian error.
However, one caveat is that  Eq.\ref{watvp} only strictly returns an unbiased estimated peculiar velocity under the assumption that the galaxy's \textit{true} peculiar velocity (not necessarily the measured peculiar velocity) is much smaller than $cz$ for that galaxy. Using the mock 2MTF surveys, \citealt{Howlett2017} explores to what extent Eq.\ref{watvp} biases the measured peculiar velocities of 2MTF. As shown in Figure 9 of \citealt{Howlett2017}, Eq.\ref{watvp} overestimates the large positive peculiar velocities, whilst underestimates the large negative peculiar velocities.


\section{Gaussianizing the peculiar velocities}\label{sec:box}

To preserve the Gaussianity and avoid any assumption on the unknown  true velocity of the galaxy compared to its redshift, we can instead Gaussianize a peculiar velocity of Eq.\ref{travp}. In this section, we will introduce the algorithm used to performing the Gaussianization. 

\subsection{Box-Cox transformation}

\citealt{Box1964} developed a power transform technique that makes the non-Gaussian distributed data more normal distribution-like \citep{Sakia1992}. The Box-Cox (BC) transformation also has been introduced into cosmology in order to offset the non-Gaussianity of measurements. For example, \citealt{wang2018} studys the BC transformation of the density power spectrum, \citealt{Qin2019b} use the BC transformation to Gaussianize the momentum power spectrum to fit the growth rate of the large scale structure. In this paper, we will apply the BC transformation to the peculiar velocities to obtain Gaussianized peculiar velocities. 

For a set of non-Gaussian distributed data $\{v_i | i=1,2,...,M\}$ ( i.e. the normalized  histogram of the data set is not Gaussian), the BC transformation of $v_i$ is defined as \citep{Box1964}
\be\label{boxc1}
Y_i\equiv\left \{
\begin{aligned}
&\frac{(v_i+\delta)^{\lambda}-1}{\lambda}, &\lambda \neq 0\\
&\ln(v_i+\delta)~, &\lambda = 0
\end{aligned}
\right.
\ee
where $\delta$ is a shift to the whole data set in order to keep  all the data being positive. Such  shift will not change the analysis of variance \citep{Box1964,Sakia1992}. $\lambda$ is the transformation parameter for the whole data set, it can be estimated by maximizing the following logarithmic likelihood function \citep{Box1964,Sakia1992,Qin2019b}
\be\label{boxclog}
L(\lambda) \sim (\lambda-1)\sum_i^M\ln(v_i+\delta)-\frac{M}{2}\ln\left( \frac{\sum_i(Y_i(\lambda)-\bar{Y}(\lambda))^2}{M}   \right),
\ee
following the steps presented in \cite{Box1964} and \cite{Qin2019b}.  
Using Eq.\ref{boxc1}, applying the estimated $(\lambda,~\delta)$ to each $v_i$ of the data set, we can obtain a set of $Y_i$, then $\{Y_i|i=1,2,...,M\}$ is the corresponding Gaussianized data set. 

\subsection{Methodology}\label{sec:box1}

The starting point is the non-Gaussian PDF of peculiar velocity, Eq.\ref{PDFv}. We will present an algorithm of applying BC transformation to Eq.\ref{PDFv}. For convenience, we assume the $P(\eta)$ term of Eq.\ref{PDFv} is a Gaussian function, i.e. 
we assume the measured log-distance ratio of the $n$-th galaxy, $\eta_n$, has Gaussian error $\epsilon_n$, then the PDF of log-distance ratio of this galaxy is given by the Gaussian equation 
\be\label{PDFeat}
P(\eta)=\frac{1}{\sqrt{2 \pi \epsilon^2_n}}\exp\left(- \frac{    (\eta-\eta_n)^2  }{ 2 \epsilon^2_n }\right)~.
\ee
 To clarify, this assumption is not necessary to the   following presented algorithm, we make this assumption  is  to clearly and conveniently present our algorithm. 
 Although $P(\eta)$ is assumed to be a Gaussian function here, due to the non-linear relation between $P(\eta)$ and $P(V)$ in Eq.\ref{PDFv}, the resultant $P(V)$ is not Gaussian. Therefore the Gaussian assumption of peculiar velocity in the past literature is not true and should be abandoned.

To clearly present the algorithm, we randomly choose one galaxy from the 2MTF catalogue, the 2MASS ID of this galaxy is `2MASX09582105+3222119', the log-distance ratio (and error), redshift and peculiar velocity (estimated using Eq.\ref{travp}) of this galaxy are listed in Table~\ref{galexamp}, we also listed its Galactic longitude $\ell$ and latitude $b$ in the table.

The BC transformation parameter $\lambda$ is estimated from a set of samples using Eq.\ref{boxclog}. For a galaxy, we first need to generate a set of samples which has PDF of Eq.\ref{PDFv}, then estimating $\lambda$ for this galaxy using these samples. The details of the algorithm are presented as follows \footnote{The code for the algorithem can be downloaded from \url{https://github.com/FeiQin-cosmologist/GaussPv}}:

\begin{table}   \centering
\caption{The properties of the 2MTF galaxy `2MASX09582105+3222119'.}
\begin{tabular}{|c|c|}
\hline
\hline
\multicolumn{2}{|c|}{2MASX09582105+3222119}\\
\hline
 $\eta$   & 0.133183   \\
\\
 $\epsilon$   & 0.108191   \\
\\
 $cz$   & 1748  km s$^{-1}$  \\
\\
 $V$ & 460.15 km s$^{-1}$ \\
\\
 $l$ & 194.22$^{\circ}$ \\
\\
 $b$ & 52.32$^{\circ}$ \\
\hline
\end{tabular}
 \label{galexamp}
\end{table}

{\bf {\noindent   (i) Generating a spline function $P_{\mathrm{spl}}(V)$ for Eq.\ref{PDFv}:} }

\begin{figure} 
\centering
 \includegraphics[width=100mm]{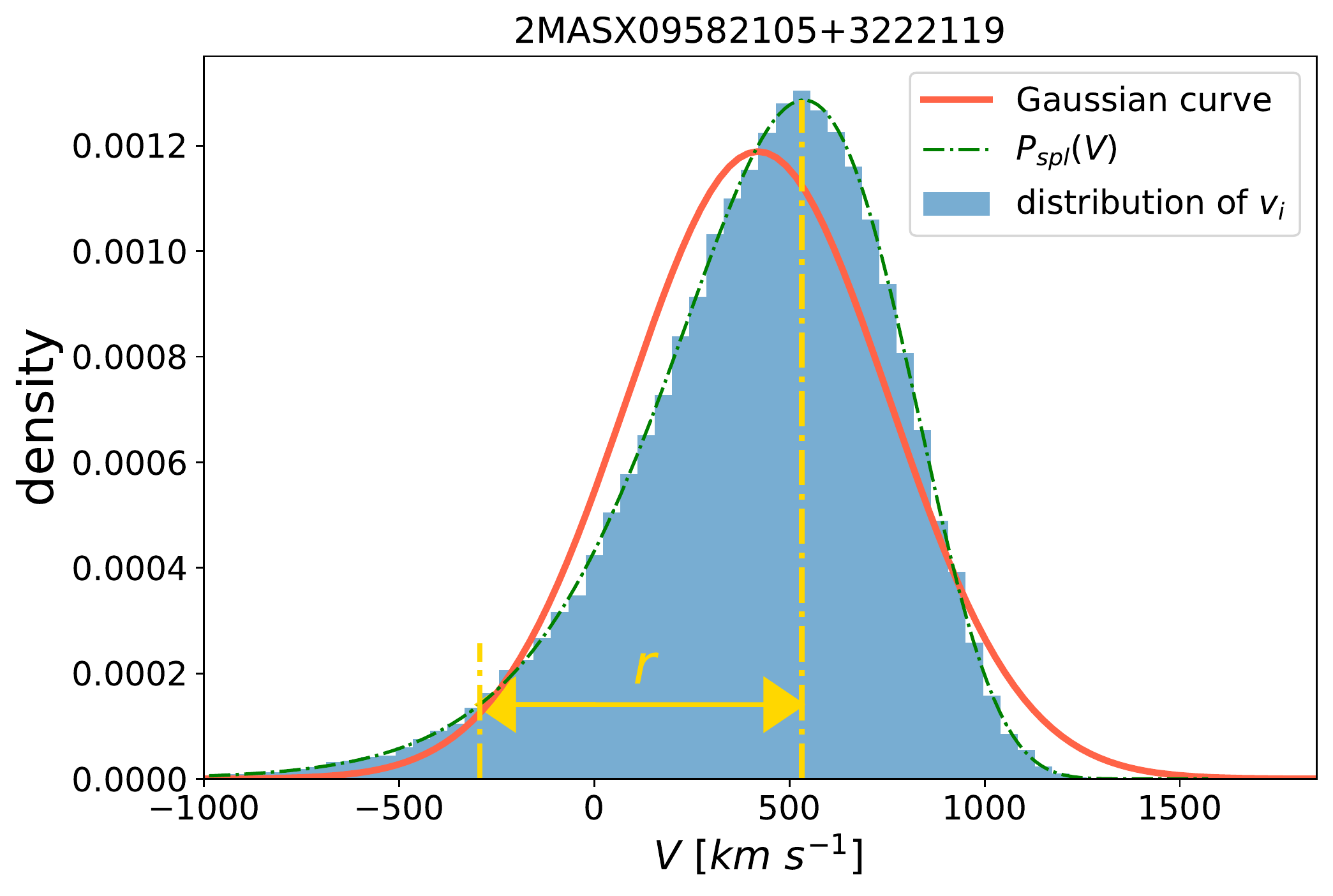}
 \caption{
 The dashed green curve is the spline function of Eq.\ref{PDFv} for `2MASX09582105+3222119'. The blue bars show the normalized histogram of the
 velocities samples $\{v_i| i=1,2,...,M\}$ generated from step (iii) according to Eq.\ref{PDFv}. For comparison, the red curve indicates the position of Gaussian PDF, centered in the mean value of these velocities samples, width calculated from the std of these velocities samples. The yellow arrow indicates the width where $P_{\mathrm{spl}}(V)=0.1\times$max$[P_{\mathrm{spl}}(V)]$.}
 \label{PVss}
\end{figure}

Generating a set of $\eta \in [-1,1] $\footnote{From Fig.\ref{logds} we know the interval $\eta \in [-1,1] $ is large enough to cover all the 2MTF galaxies.}, calculating $P(\eta)$ from Eq.\ref{PDFeat}, where $\eta_n=0.133183$ and $\epsilon_n=0.108191$. 
Calculating the corresponding  $z_h$ and $d_h$ using the $\eta$ values and $cz=$1748  km s$^{-1}$. Then 
using Eq.\ref{travp} and Eq.\ref{PDFv} to calculate a set of velocities $v$ and the corresponding $P(v)$. Then we obtain a set of interpolation points $q=[v,~P(v)]$ and the corresponding spline function $P_{\mathrm{spl}}(V)$,  as shown in the dashed green curve in Fig.\ref{PVss}. The curve is bias from Gaussian significantly, i.e. the peculiar velocity of this galaxy does not have Gaussian error.

{\bf {\noindent   (ii) Generating a spline function for the inverse cumulative function corresponding to Eq.\ref{PDFv}:} }

\begin{figure} 
\centering
 \includegraphics[width=100mm]{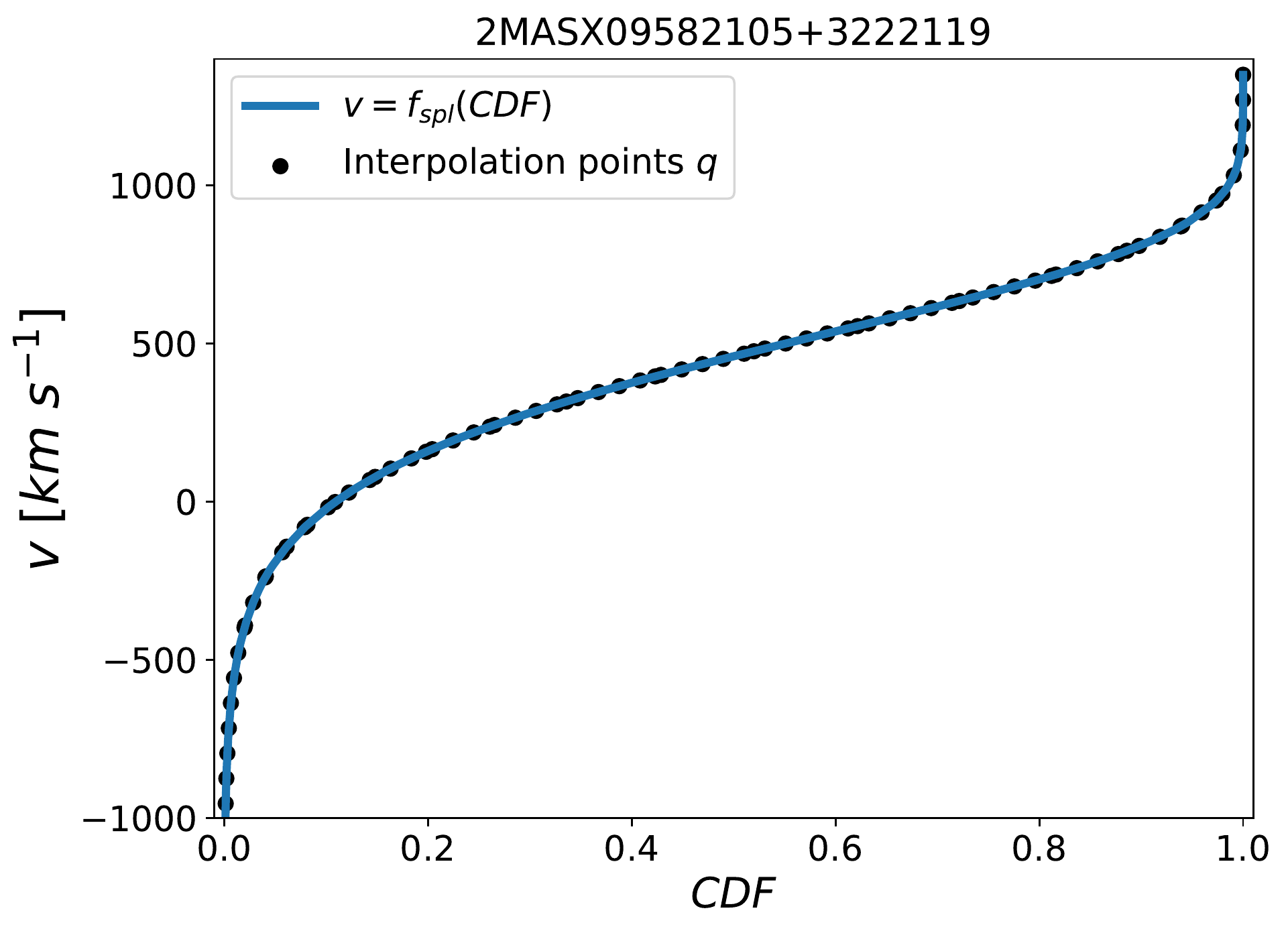}
 \caption{The black dots are the interpolation points generated in step (i). The blue curve shows the inverse spline function of CDF for `2MASX09582105+3222119'.}
 \label{CDFvs}
\end{figure}

Using the above interpolation points $q$ and the spline function $P_{\mathrm{spl}}(V)$, one can numerically estimate the cumulative distribution function (CDF) using
\be 
CDF=\int_{\mathrm{min}(v)}^{v} P_{\mathrm{spl}}(v') dv'~.
\ee
Then we can obtain a spline function $V=f_{\mathrm{spl}}(CDF)$, which is the inverse function of CDF. As shown in Fig.\ref{CDFvs}.

{\bf {\noindent   (iii) Generating velocity samples which has a PDF of Eq.\ref{PDFv}:} }

Generate $M=150000$ uniform distributed random points in the interval of $[0,1]$ as the input to $V=f_{\mathrm{spl}}(CDF)$ to obtain velocity samples $\{v_i| i=1,2,...,M\}$. In Fig.\ref{PVss}, the blue bars shows the normalized histogram  of these samples, which matches the $P_{\mathrm{spl}}(V)$ curve.

{\bf {\noindent   (iv) BC transform of the velocity samples:} }

  Setting $\delta=25r$, where $r$ corresponding to the width where $P_{\mathrm{spl}}(V)=0.1\times$max$[P_{\mathrm{spl}}(V)]$, as shown in the yellow arrow in Fig.\ref{PVss} (See Section \ref{vdfgb} and Appendix \ref{AP3} for more discussion about the choice of $\delta$).  
 
\begin{figure} 
\centering
 \includegraphics[width=100mm]{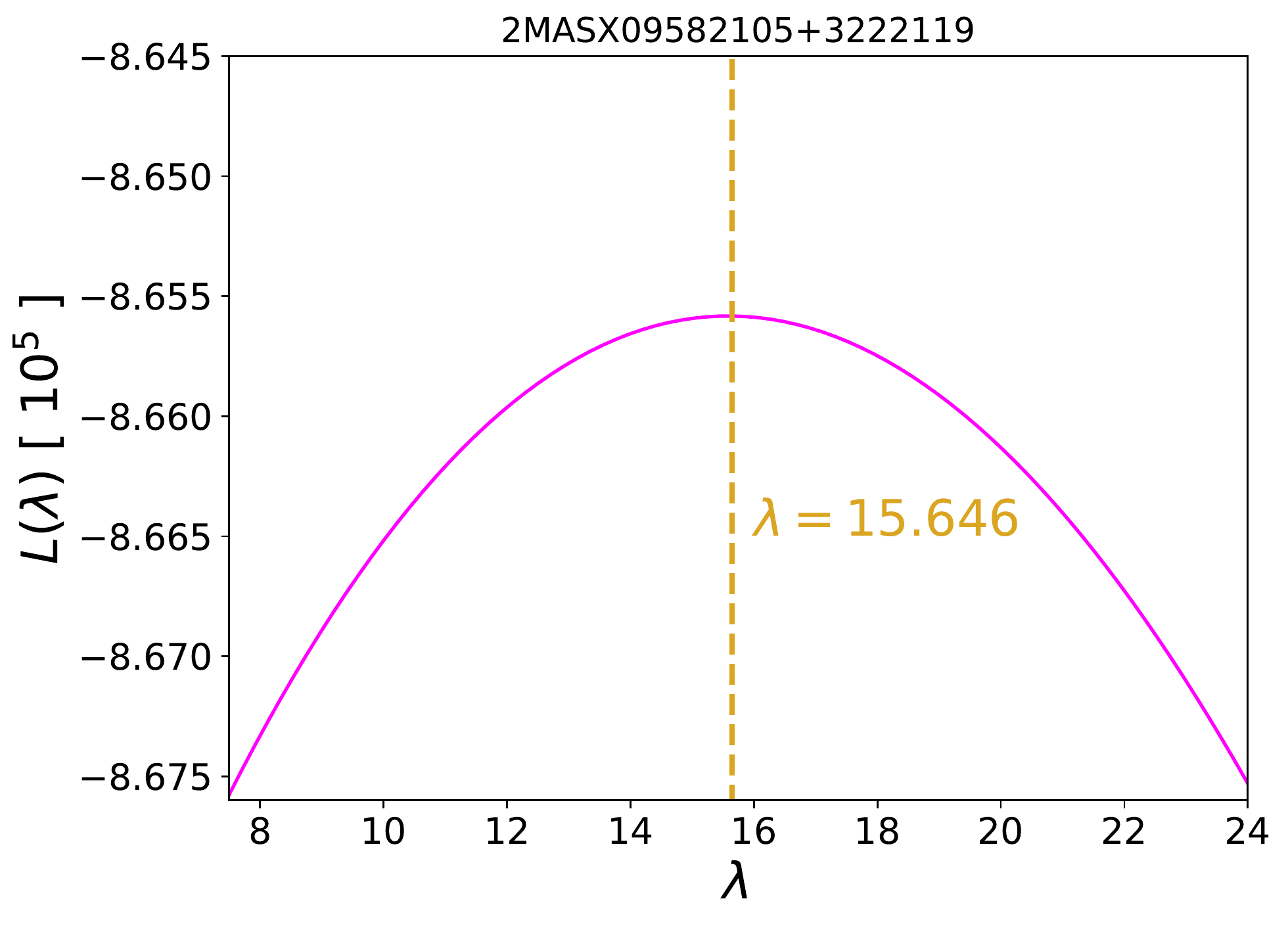}
 \caption{The pink curve shows   the logarithmic likelihoods $L(\lambda)$ for `2MASX09582105+3222119'. The yellow dashed vertical line indicates the position of $\lambda$ which maximizes $L(\lambda)$.}
 \label{BCLBDs}
\end{figure}

 Choosing a $\lambda$ in $(-\infty,+\infty)$\footnote{In practice, the estimated $\lambda$ is all in $[5,~30]$ for 2MTF.}.  Substituting $\lambda$, $\delta$ and $v_i$ into Eq.\ref{boxc1}, then into Eq.~\ref{boxclog} to calculate a $L(\lambda)$. Repeat this step to find the value of $\lambda$ that maximizes $L(\lambda)$. As shown in Fig.\ref{BCLBDs}, $\lambda=15.646$ is the best estimated BC transformation parameter for this galaxy.

 {\bf {\noindent   (v) Gaussianize the peculiar velocity:} }

\begin{figure} 
\centering
 \includegraphics[width=100mm]{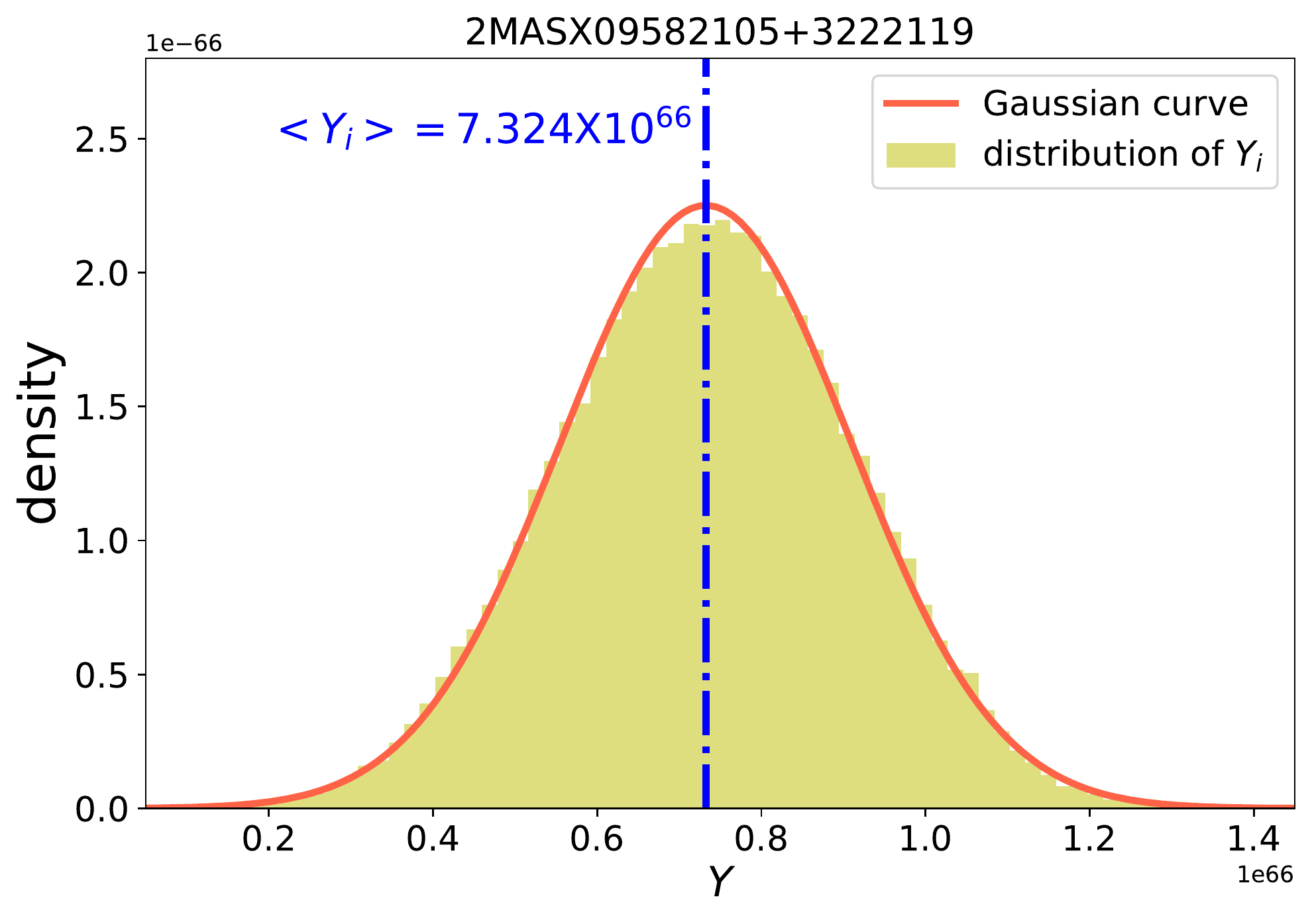}
 \caption{The yellow bars show the normalized histogram of the BC transformed `velocity' samples $\{Y_i| i=1,2,...,M\}$ from step (v). For  comparison,  the  red  curve  indicates  the position of Gaussian PDF, centered in the mean value of these samples, width calculated from the std of these samples. The blue dashed vertical line indicates the position of the mean of these samples, the mean is $7.324\times10^{66}$.}
 \label{BCys}
\end{figure}

   Plugging the galaxy's peculiar velocity $V=$460.15 km s$^{-1}$ and  $\lambda=15.646$ as well as the above $\delta$ into  Eq.\ref{boxc1}, we obtain the $Y=$7.3$\times$10$^{66}$. This is the BC transformed `velocity' for this galaxy.
   Substituting $\lambda=15.646$, $\delta$ and $\{v_i| i=1,2,...,M\}$ into Eq.\ref{boxc1} to obtain a data set 
   $\{Y_i| i=1,2,...,M\}$.
   The standard deviation (std) of this data set, $\sigma=$1.77$\times$10$^{65}$ is measurement error of $Y$. 
   In fact, the mean value of $\{Y_i| i=1,2,...,M\}$, $<Y_i>$=$7.3 \times$10$^{66}$ is equal to the $Y$ transformed directly from $V=$460.15 km s$^{-1}$. 
   As shown in Fig.\ref{BCys}, the distribution of $\{Y_i| i=1,2,...,M\}$ (yellow bars) matches the Gaussian curve (red curve) very well.

Applying the above algorithm to each of the 2MTF galaxies, we finally obtain the $(\lambda,\delta,Y,\sigma)$ for each galaxy. The data set of $(Y, \sigma)$ are the Gaussianized `velocities' which we can use to measure power spectrum, two-point correlation and cosmic flows. When fitting the measurements to the model, we also need to apply the same $(\lambda,\delta)$ to the modeled velocity value for each galaxy. 

To reiterate, the above algorithm is independent from $P(\eta)$.
Although  $P(\eta)$ is assumed to be Gaussian in Eq.\ref{PDFeat}, however, this assumption is not necessary to the above algorithm. $P(\eta)$, in principle, can be chosen as any distribution. We even do not need to know the analytic expression of $P(\eta)$, as long as we can obtain an numeral function for $P(\eta)$, we can still calculate $P_{\mathrm{spl}}(V)$ in step (i), and then Gaussianize the peculiar velocity use the above algorithm. This shows the flexibility of the  above algorithm.

\subsection{$\lambda$ as a function of $\epsilon$}\label{vdfgb}

\begin{figure} 
\centering
 \includegraphics[width=100mm]{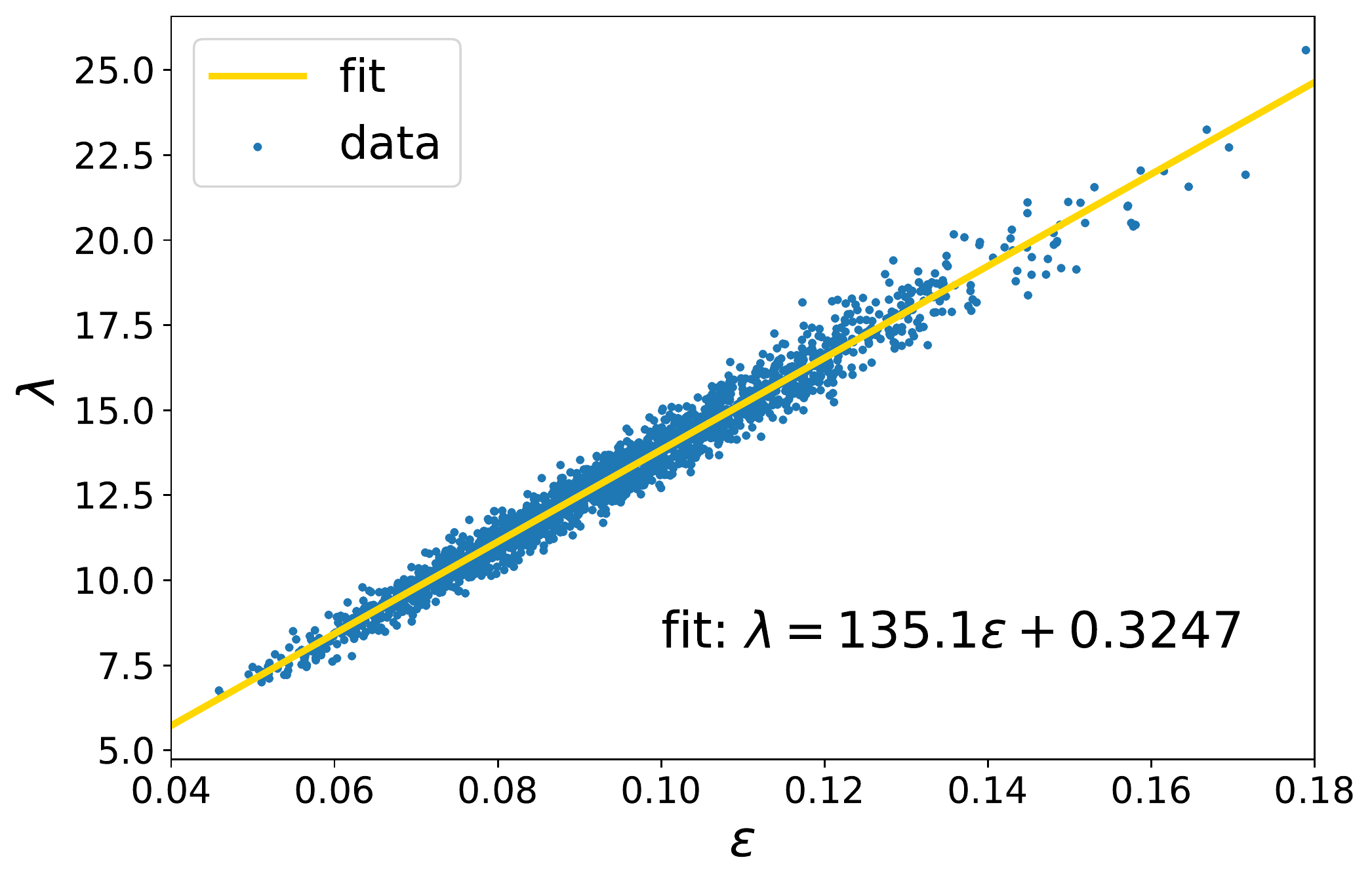}
 \caption{$\lambda$ as a function of $\epsilon$ for all 2MTF galaxies. The yellow line is the best fit to the blue data points.}
 \label{BCyss}
\end{figure}

In Fig.\ref{BCyss}, we plot the BC transformation parameter $\lambda$ against the measurement error of log-distance ratio  $\epsilon$ for all the 2MTF galaxies, the yellow line is the best fit to the dots, and the fit equation is given by:
\be \label{lbdep} 
\lambda= k \epsilon+b~,
\ee
For $\delta=25r$, the fit result is:
\be  \label{asw}
k=135.1~,~~b=0.3247~,
\ee
The relation Eq.\ref{lbdep} is also existed for the 2MTF mocks, see Appendix \ref{AP2} and Fig.\ref{BCyssA} for more discussion. 
Does this relation also exist for other surveys? From the details of the algorithm presented in Section \ref{sec:box1}, 
we find that the BC transformation parameter $\lambda$
does not depend on any particular survey. In Appendix \ref{AP2}, we find this relation also existed in the 6dFGSv surveys. 
6dFGSv \citep{Springob2014} is a Fundamental Plane survey. Therefore, the linear relation Eq.\ref{lbdep} exists for both Tully-Fisher and Fundamental Plane surveys (at least, for 2MTF and 6dFGSv). 
In the future work, we also need to explore whether this relation is true for the peculiar velocity samples measured from Type-Ia supernovae (as well as any other distance measurements techniques). If this relation widely exists, then one can compute 
$\lambda$ directly from Eq.\ref{lbdep}, rather than performing the whole algorithm. This will be very time saving for plenty of mocks and the upcoming larger surveys, such as SkyMapper \citep{2018arXiv180107834W}, DESI \citep{DESI2016}, LSST \citep{2008arXiv0805.2366I},  WALLABY \citep{2012PASA...29..359K,Koribalski2020} and Taipan Galaxy Survey \citep{2017PASA...34...47D}, .

The choice of $\delta$ can change the fit parameter $k$, but will not change the linear relation and $b$. Choosing smaller  $\delta$ will result in smaller $k$, but may not  keep all the samples positive in Eq.\ref{boxc1}. Too large $\delta$ will result in overflow in the memory of computer.  See Appendix \ref{AP3} for more discussion.

\section{Bulk flow estimation techniques} \label{sec:BKest}

In this paper, we take the bulk flow measurement as an example to show the implementation of the Gaussianized peculiar velocities. We will introduce a new bulk flow estimator that estimates bulk flow from the Gaussianized peculiar velocities. We also using mocks to test the technique and compare it to the  techniques used in other literature. 
In the future work, we will extend to explore new techniques for the power spectrum and two-point correlation measurements as well as the density field reconstruction using the Gaussianized peculiar velocities.

 \subsection{Bulk flow estimation techniques in previous work}

Galaxies' peculiar motions form the cosmic flow field in the nearby Universe. The bulk flow velocity is the dipole component of the cosmic flow field \citep{Lister1989,Jaffe1995,Parnovsky2001,Feldman2010,Qin2019a}. Measuring the bulk flow velocity and compare to the cosmological model prediction enables us to test whether the model accurately describe the motion of galaxies in the nearby Universe. In the past literature, bulk flow is usually measured in velocity space and log-distance ratio space. 

Firstly, measuring the bulk flow in velocity space (v-space). The two main v-space measurement techniques are  maximum likelihood estimation (MLE, \citealt{Kaiser1988}), and minimum variance (MV) estimation \citep{Watkins2009,Feldman2010}. In this paper we only focus on the MLE technique. In \citealt{Kaiser1988}, under the assumption that peculiar velocities have Gaussian errors,  the likelihood of $N$ line-of-sight peculiar velocities can be written as
\be\label{tramlesa}
S({\bf B},\alpha_{\star})=\prod_{n=1}^{N}\frac{1}{\sqrt{2 \pi \left(\alpha^2_n+\alpha_{\star}^2 \right)}}\exp\left(-\frac{1}{2}  \frac{    (V_{n}-{\bf B} \cdot {\bf \hat{r}}_n)^2  }{  \alpha^2_n+\alpha_{\star}^2 }\right)~,
\ee
where the vector ${\bf B}=[B_x,B_y,B_z]$ is the bulk flow velocity to be estimated, ${\bf \hat{r}}_n$ is the unit vector point to the $n-$th galaxy, $\alpha_n$ is the measurement error of $V_n$. Finally  $\sigma_{\star}$ is introduced to account for the intrinsic scatter of the peculiar velocities, and is usually assumed to be 300 km s$^{-1}$ \citep{Sarkar2007,Scrimgeour2016} (though in this paper we vary it as a free parameter).
In this technique,the peculiar velocities estimated from Eq.\ref{watvp} can be the input to Eq.\ref{tramlesa}. For convenience, we call this technique $w$MLE.

Secondly, to avoid the non-Gaussianity of the estimated peculiar velocities, the measurement of bulk flow can be performed in the log-distance ratio-space, or $\eta$-space. \citealt{Nusser1995,Nusser2011,Qin2018,Qin2019a}  used the so-called $\eta$MLE to measure the bulk flow. \citealt{Boruah2020} measures the bulk flow in distance modulus space, since $\eta$ is simply linearly converted from the distance modulus, their method and any other similar method also be classified as $\eta$MLE. In this paper we use the $\eta$MLE of \citealt{Qin2018}, and the algorithm is clearly presented in Section 4.2 of that previous work.

\subsection{Bulk flow estimator}\label{ijnh}

In this section, we will introduce the bulk flow estimator that estimates bulk flow from the BC-transformed peculiar velocities. 

After we obtain the $(\lambda,\delta,Y,\sigma)$ for each galaxy, we can write the likelihood of $N$ galaxies, each with $(Y_n,\sigma_n)$, as
\be\label{smle}
L({\bf B},\sigma_{\star})=\prod_{n=1}^{N}\frac{1}{\sqrt{2 \pi \left(\sigma^2_n+\sigma_{\star}^2 \right)}}\exp\left(-\frac{1}{2}  \frac{    (Y_{n}-F_n({\bf B}))^2  }{  \sigma^2_n+\sigma_{\star}^2 }\right)~,
\ee
where:
\be
F_n({\bf B})=\left \{
\begin{aligned}
&\frac{({\bf B} \cdot {\bf \hat{r}}_n+\delta_n)^{\lambda_n}-1}{\lambda_n}, &\lambda_n \neq 0\\
&\ln({\bf B} \cdot {\bf \hat{r}}_n+\delta_n)~, &\lambda_n = 0
\end{aligned}
\right.
\ee 
where the vector ${\bf B}$ is the bulk flow velocity to be estimated, ${\bf \hat{r}}_n$ is the unit vector point to the $n$-th galaxy. $\sigma_{\star}$ is introduced to account for the intrinsic scatter of the velocities. In this paper, we set it as a free parameter. $F_n({\bf B})$ is the BC transformation of the model peculiar velocity for the $n$-th galaxy, transformed using $(\lambda_n,\delta_n)$ of that galaxy.

The maximum likelihood ${\bf B}$ cannot be obtained analytically due to the non-linear relationship between the model ${\bf B}$ and $F_n({\bf B})$. Instead, combining uniform priors on the $\sigma_{\star}$ and ${\bf B}$ with the likelihood in Eq.~\ref{smle} to obtain the posterior probability of these four independent parameters, we can estimate the bulk flow using the Metropolis-Hastings Markov chain Monte Carlo (MCMC) algorithm. The flat priors of the four parameters are in the interval $B_{i}\in[-1200,+1200]$ km s$^{-1}$  and $\sigma_{\star}\in[-1000,+1000]$ h km s$^{-1}$ Mpc $^{-1}$.

The measurement error of the bulk flow component,  $e_{B_i}$ ($i=x,y,z$) is the std of the MCMC samples of the corresponding MCMC chain\footnote{We use the {\tiny PYTHON } package {\tiny emcee } \citep{Foreman-Mackey2013} to perform the MCMC. For each of the four parameters, we use 24 walkers, and for each walker, we generated 100,000 MCMC samples. Therefore, there are 2,400,000 samples in each of the four MCMC chains. This is smooth fairly enough to estimate the measurements errors.}. The measurement error of the bulk flow amplitude, $e_{B}$ is calculated using \citep{Scrimgeour2016,Qin2018}:
\be\label{bke2}
e^2_B=JC_{ij}J^T~,~~(i=1,2,3)~,
\ee
where $J$ is the Jacobian of the bulk flow, $\partial B/\partial B_i$. $C_{ij}$ is the covariance of the bulk flow components calculated using the MCMC samples.

In the following section,
we will test and compare the above bulk flow estimation technique to $w$MLE and $\eta$MLE using mock 2MTF surveys.

\subsection{Testing using mocks}

In order to compare and test how well the bulk flow estimators are expected to recover the true bulk flow from the 2MTF survey, we applied the three estimators to 16 mock 2MTF catalogues.

The `true' bulk flow velocity, ${\bf B}_t$ within each mock is defined by averaging over the true galaxy velocities ${\bf v}_t$ along orthogonal axes \citep{Qin2018,Qin2019a}
\be
B_{t,i}=\frac{1}{N}\sum^N_{n=1}v_{t,in}~,~~(i=x,y,z)~,
\ee
where ${\bf v}_t$ is know from the simulations. Only the mock 2MTF galaxies in the simulation are used to compute ${\bf B}_t$.

\begin{figure}  
\centering
  \includegraphics[width=100mm]{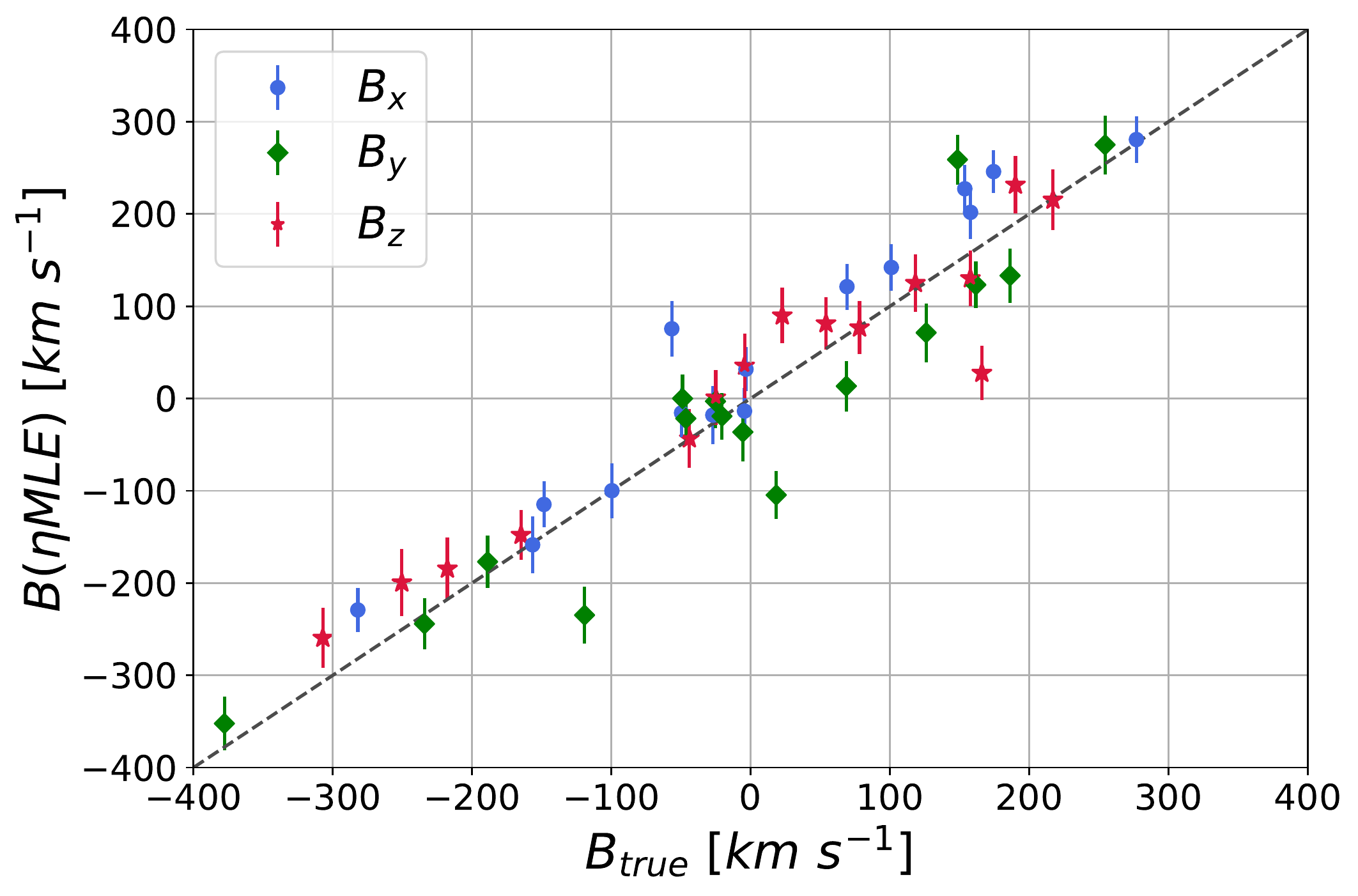}
     \includegraphics[width=100mm]{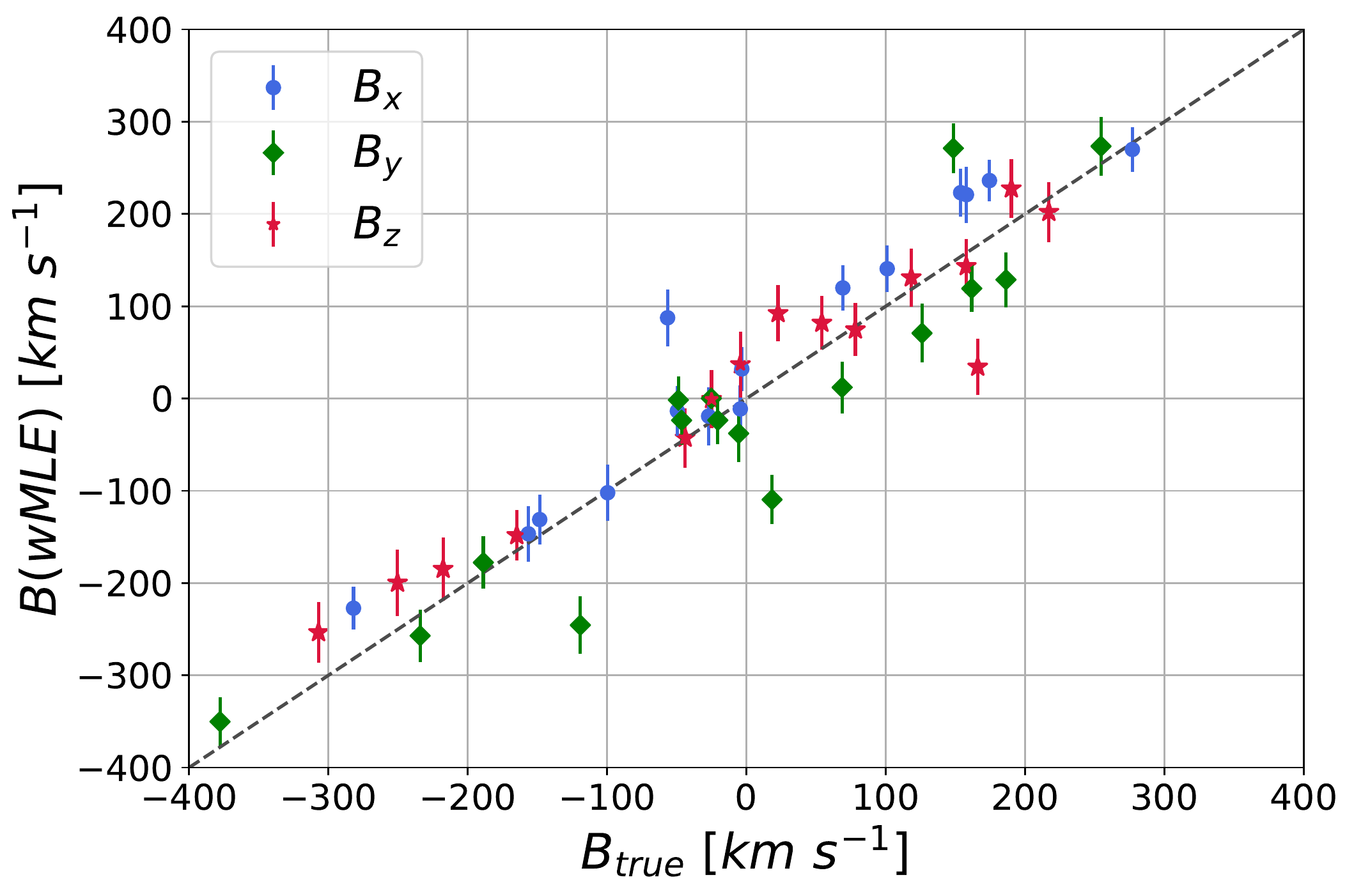}
    \includegraphics[width=100mm]{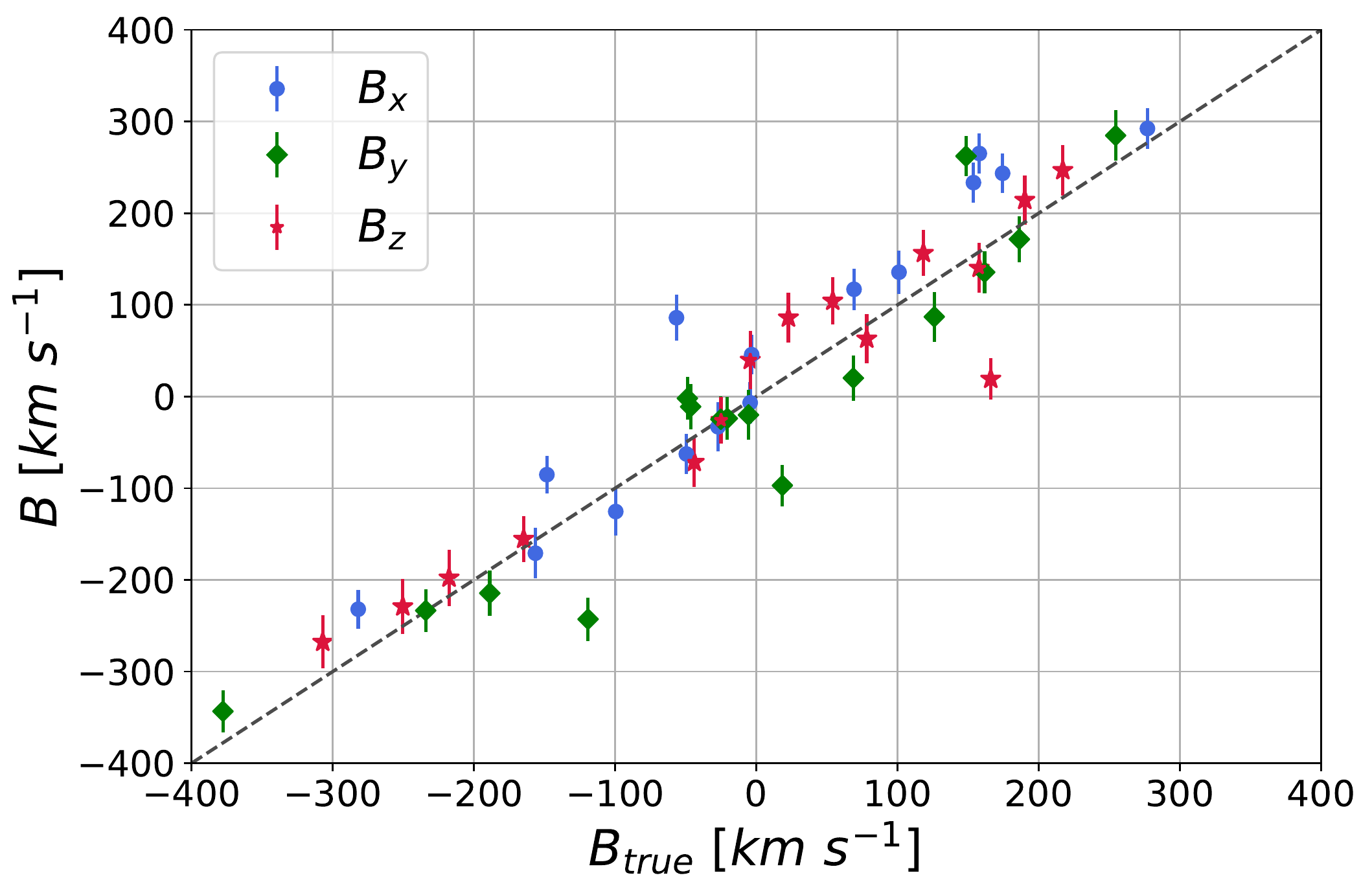}
 \caption{  Comparing the measured bulk flow for the 16 2MTF mocks in equatorial coordinates. The top and middle panels are for the $\eta$MLE and $w$MLE, respectively. The bottom panel is  measured from the Gaussianized velocities along with the estimator in Section \ref{ijnh}.}
\label{2mtfmock}
\end{figure}

Fig.\ref{2mtfmock} shows the measured bulk flow against the true bulk flow in equatorial coordinates. All three bulk flow estimators can recover the true bulk flow. The top panel is the $\eta$MLE measurement. 
The middle panel is the $w$MLE measurement. 
The bottom panel is the measurement using the technique of Section \ref{ijnh}. 
 The scatters of the points in the three panels are  most likely due to the intrinsic scatter $\sigma_{\star}$ (or $\alpha_{\star}$ of wMLE)  of the true velocities in the mocks \citep{Qin2018,Qin2019a}. $\sigma_{\star}$  (or $\alpha_{\star}$) accounts for the non-linear peculiar motions of galaxies. In the future, a more accurate peculiar velocity estimator, which can more accurately predict and model for non-linear motion, needs to be developed.  Such new estimators will result in more non-Gaussianity than Eq.\ref{travp}, indicating the importance and usefulness of our method for Gaussianizing the estimated peculiar velocities.

\begin{figure}  
\centering
  \includegraphics[width=100mm]{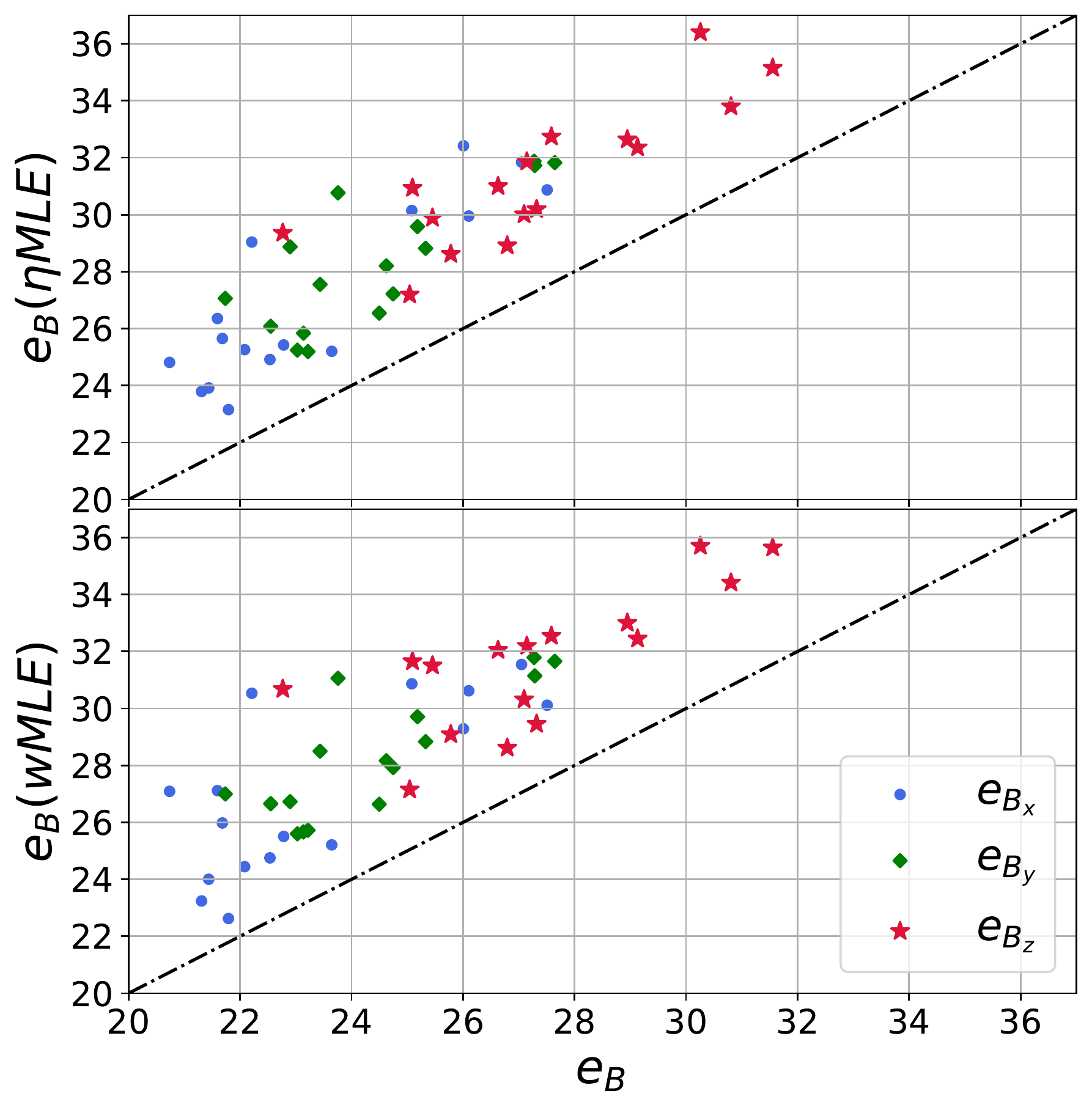}
 \caption{ Comparing the measurement errors of bulk flows measured from the Gaussianized peculiar velocities (of 16 2MTF mocks in equatorial coordinates) to those of $\eta$MLE (top panel) and $w$MLE (bottom panel). 
 }
\label{2mtfmockerr}
\end{figure}

If only having a look at the scatters of the symbols in the three panels (the difference between the measured values and the true values), one can find that the scatters are almost the same for the three estimators. However,  the technique of Section \ref{ijnh} gives smaller measurements errors.
We plot the measurement errors of the bulk flows given by this technique against those of $\eta$MLE and $w$MLE, as shown in Fig,\ref{2mtfmockerr}, we find this technique gives smaller measurement errors comparing to both $\eta$MLE and $w$MLE.

  Using the Gaussianized peculiar velocity fields to measure the bulk flow can reduce the measurements errors.
    In the future, we will also explore whether using the Gaussianized peculiar velocities can  reduce the measurement errors of the power spectrum and two-points correlation, and then reduce the measurement errors of the cosmological parameters.  

The value of $\delta$ will not change the bulk flow measurements. See Appendix \ref{AP3} for more discussion.

\section{Bulk flow Result and discussion}\label{sec:dis}

\subsection{Results}

\begin{figure}  
\centering
  \includegraphics[width=120mm]{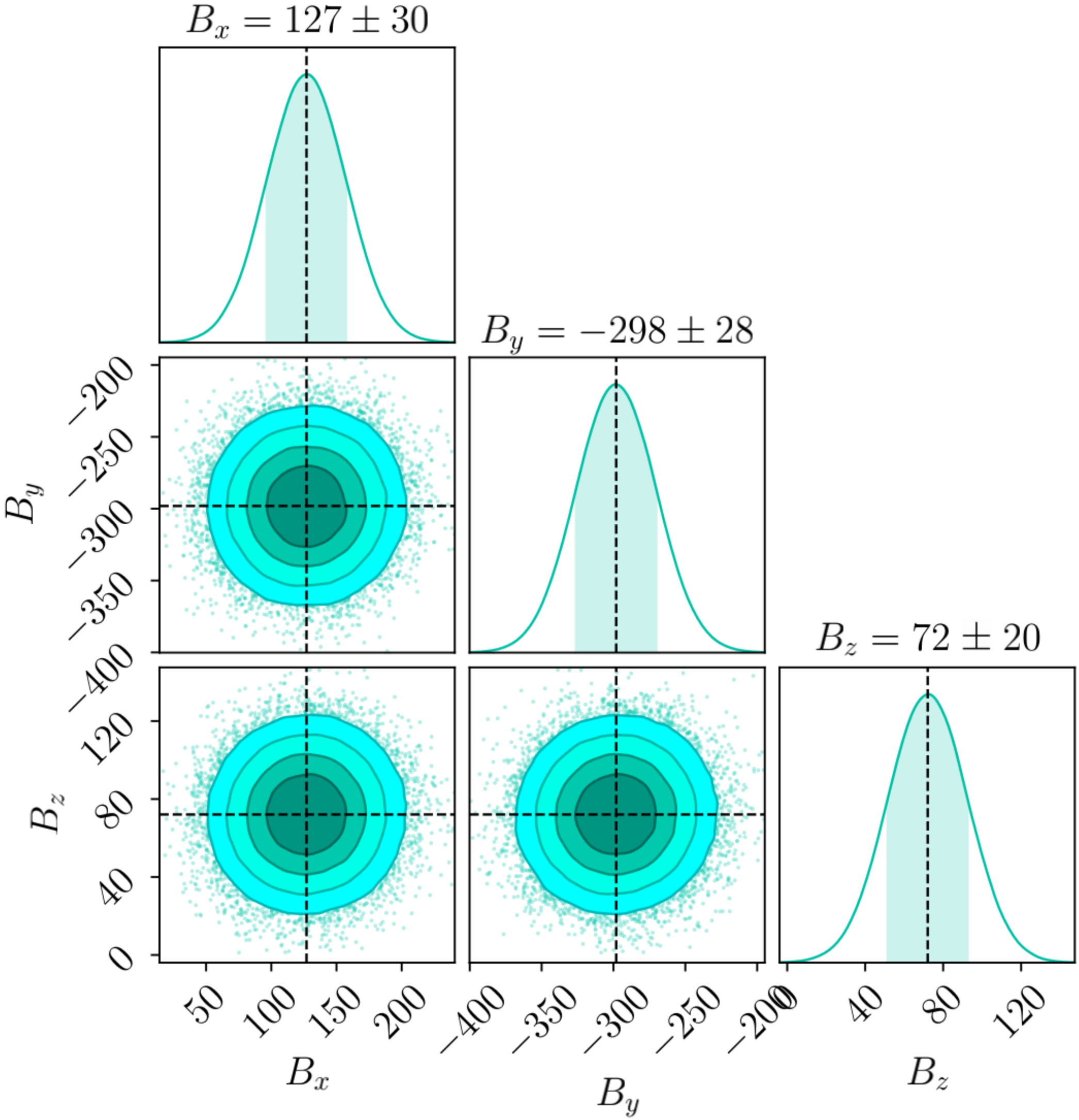}
 \caption{The bulk flow velocity measurements of the 2MTF survey. The marginalised histograms shows the distribution of the MCMC samples of $B_x$, $B_y$ and $B_z$. The shaded region in the histograms is $1\sigma$. The filled 2D contours indicate the 1, 1.5, 2 and 2.5$\sigma$ regions. The vertical dashed line indicates the best estimated value of $B_x$, $B_y$ and $B_z$.}
\label{2mtfbkfmc}
\end{figure}

 Fig.\ref{2mtfbkfmc} shows the bulk flow velocity measurement in Galactic coordinates
using the Gaussianized 2MTF peculiar velocities along with the bulk flow estimator in Section \ref{ijnh}.  The vertical dashed line indicates the best estimated bulk flow velocity components $B_i$ $(i=x,y,z)$. The histograms shows the distribution of MCMC samples for each $B_i$,  
the shaded regions are the $1\sigma$ measurement errors of $B_i$.

For comparison, in Table~\ref{bkflb}, we list the measured bulk flow velocity and its direction using the three estimators. Our new estimator gives the smallest error compare to $\eta$MLE and $w$MLE.

\begin{table*}   \small
\caption{Comparing the bulk flow velocities of 2MTF survey measured from the three estimators.  
}
\begin{tabular}{|c|c|c|c|c|c|c|c|}
\hline
\hline

&  $|{\bf{B}}|$ & $B_x$ &   $B_y$  &  $B_z$ 
& $\ell$ & b & Depth \\

&km s$^{-1}$ &   km s$^{-1}$  & km s$^{-1}$&km s$^{-1}$
& degree&degree &Mpc h$^{-1}$\\
\hline

  This paper  &$332.41\pm27.45$    & $126.77\pm30.18$ &   $-298.66\pm27.68$ &  $72.31\pm20.27$ &$292.99\pm 5.30$&   $12.56\pm 3.54$  &30\\
\hline

  $\eta$MLE  &$ 333.99\pm30.31$    & $116.66\pm33.81$ &   $-304.29\pm30.84$ &  $73.16\pm23.33$ &$290.98\pm 5.90$&   $12.65\pm 3.98$  &30\\
\hline

  $w$MLE  &$338.91\pm31.18$    & $107.67\pm34.45$ &   $-313.15\pm32.23$ &  $72.15\pm24.47$ &$288.97\pm 6.01$&   $12.29\pm 4.11$  &30\\
\hline
          
\end{tabular}
 \label{bkflb}
\end{table*} 

\subsection{Comparison with $\Lambda$CDM theory}

In this section, we compare the estimated bulk flow amplitude, $|{\bf{B}}|$ to the predictions from $\Lambda$CDM. At redshift zero, assuming the $\Lambda$CDM model, the growth rate $f=\Omega_m^{0.55}$ \citep{Linder2007}. The variance of the bulk flow velocity is \citep{Gorski1988,Li2012,Hong2014,Andersen2016,Qin2018,Qin2019a}
\be\label{sigB}
\sigma_B^2=\frac{H_0^2f^2}{2\pi^2}\int\mathcal{W}^2(k)\mathcal{P}(k)dk~,
\ee
where $\mathcal{P}(k)$ is the linear matter density power spectrum generate using the {\tiny CAMB} package \citep{Lewis:1999bs, 2012JCAP...04..027H},  $\mathcal{W}(k)$ is the Fourier transform of the survey window function. The computation of the accurate $\mathcal{W}(k)$ of 2MTF is clearly presented in Section 6.2 of \citealt{Qin2018}.

The PDF of the bulk flow amplitude is given by \citep{Li2012,Hong2014,Scrimgeour2016,Qin2018,Qin2019a,Boruah2020}
\be\label{pbb}
p(|{\bf{B}}|)=\sqrt{\frac{2}{\pi}}\left(\frac{3}{\sigma_B^2}\right)^{1.5}|{\bf{B}}|^2\exp\left(-\frac{3|{\bf{B}}|^2}{2\sigma_B^2}\right)~,
\ee
where the most likely $|{\bf{B}}|$ is expressed as $B_p=\sqrt{2/3}\sigma_B$, and the cosmic variance of $|{\bf{B}}|$ 
is given by $B^{~+0.419\sigma_B}_{p~-0.356\sigma_B}$ ($1\sigma$,   \citealt{Scrimgeour2016,Qin2018}).
The upper and lower limits mean that the integral of Eq.\ref{pbb} in the interval $\left[ B_p-0.356\sigma_B,  B_p+0.419\sigma_B   \right]$ is 0.68.

The $\Lambda$CDM model predicted  bulk flow amplitude for the 2MTF
is quoted from \citealt{Qin2018} and is listed in Table~\ref{bkvst}. The measurement is consistent with $\Lambda$CDM prediction.

\begin{table}   \centering
\caption{Comparing the measured bulk flow to the $\Lambda$CDM prediction. Errors on the $\Lambda$CDM prediction denote the cosmic variance.}
\begin{tabular}{|c|c|c|}
\hline
\hline
Data set &$\eta$MLE&$\Lambda $CDM\\
  &km s$^{-1}$&km s$^{-1}$\\
\hline

2MTF     & $332\pm27$ & 315$^{+161}_{-137}$ \\
\hline
\end{tabular}
 \label{bkvst}
\end{table}

\section{Conclusions}\label{conc}

We developed an algorithm that can Gaussianize the line-of-sight peculiar velocities estimated from Eq.\ref{travp}. We also find that the BC 
transformation parameter $\lambda$ is a linear function of the
the measurement error of log-distance ratio $\epsilon$. 
This relation exists for the Tully-Fisher survey, 2MTF and the Fundamental Plane survey, 6dFGSv. However, more works need to be done in the future to further exam this relation using different surveys.

We developed a bulk flow estimation technique to measure the bulk flow from the Gaussianized peculiar velocities.  We also test the estimator using 2MTF mocks, and find that measuring bulk flow from the Gaussianized peculiar velocities can reduce the measurement errors compare to $w$MLE and $\eta$MLE. In the future, we will also develop new techniques for the power spectrum and two-point correlation measurements using the Gaussianized peculiar velocities.

We have measured the bulk flow velocity using the Gaussianized 2MTF surveys. The estimated bulk flow is $332\pm27$ km s$^{-1}$ at a depth of $30h^{-1}$ Mpc, the result is consistent with the $\Lambda$CDM prediction.

\section*{Acknowledgements}
We like to thank David Parkinson for multiple discussions.

FQ is supported by the project \begin{CJK}{UTF8}{mj}우주거대구조를 이용한 암흑우주 연구\end{CJK} (``Understanding Dark Universe Using Large Scale Structure of the Universe''), funded by the Ministry of Science. 

This research has made use of the 
{\verb emcee } package \citep{Foreman-Mackey2013},
{\verb ChainConsumer } package \citep{ChainConsumer},
 {\verb SCIPY } package \citep{Virtanen2020} and  {\verb MATPLOTLIB }
package \citep{Hunter2007}.

\appendix                  

\section{$\lambda$ as a function of $\epsilon$ for 2mtf MOCKS AND 6\lowercase{d}FGS\lowercase{v} SURVEY}\label{AP2}

\begin{figure} 
 \centering
 \includegraphics[width=100mm]{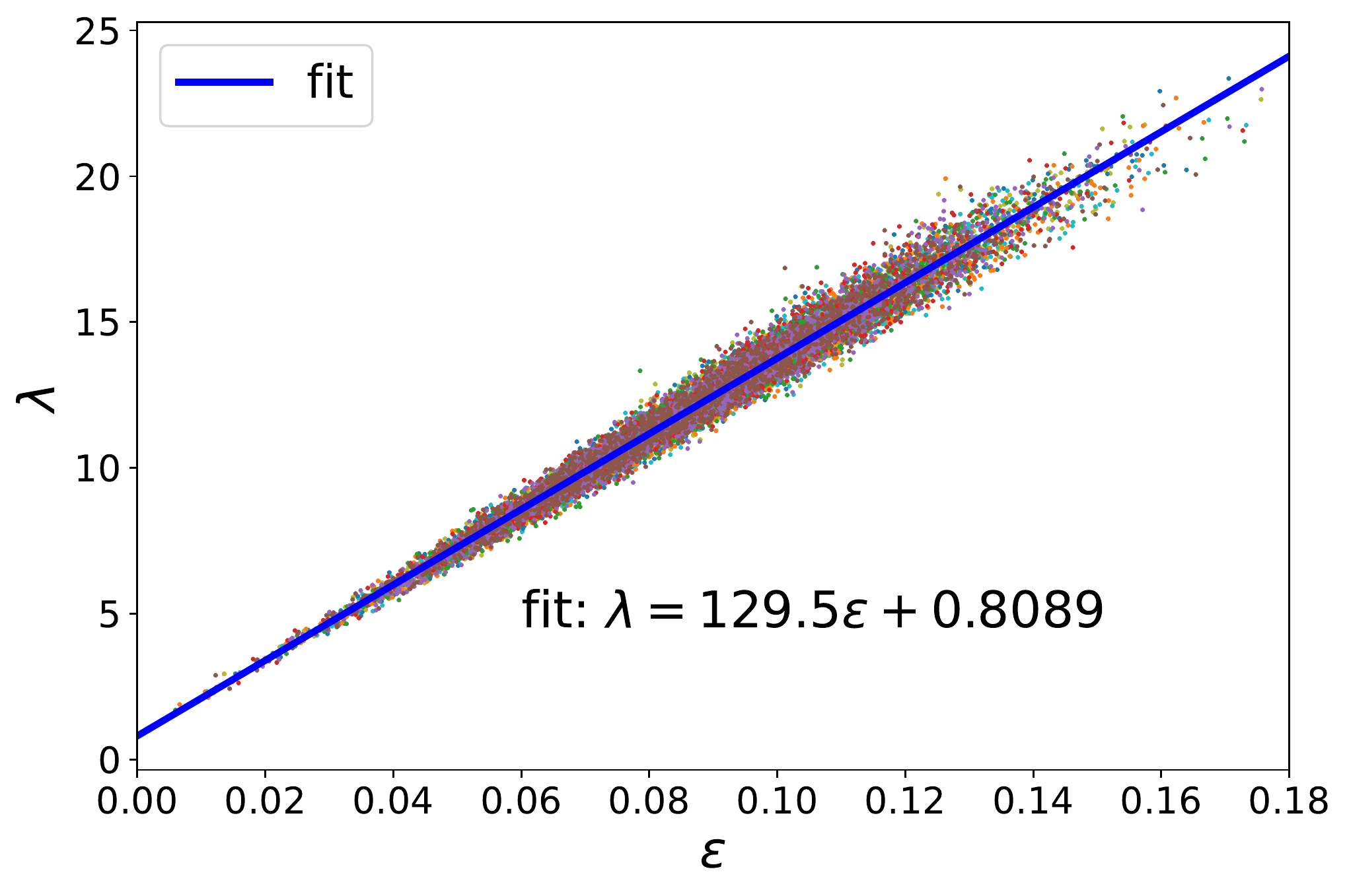}
 \caption{Same as Fig. \ref{BCyss}, but for 16 2MTF mocks. The blue curve is the best fit to the data points. The different colors of the points indicates the points from 16 different mocks.}
 \label{BCyssA}
\end{figure}

Fig.\ref{BCyssA} shows the relation between $\lambda$ and $\epsilon$ for 16 2MTF mocks.
The fit is produce by put all the 16 mocks together, the result is:
\be  
k=129.5~,~~b=0.8089~.
\ee

\begin{figure} 
 \centering
 \includegraphics[width=100mm]{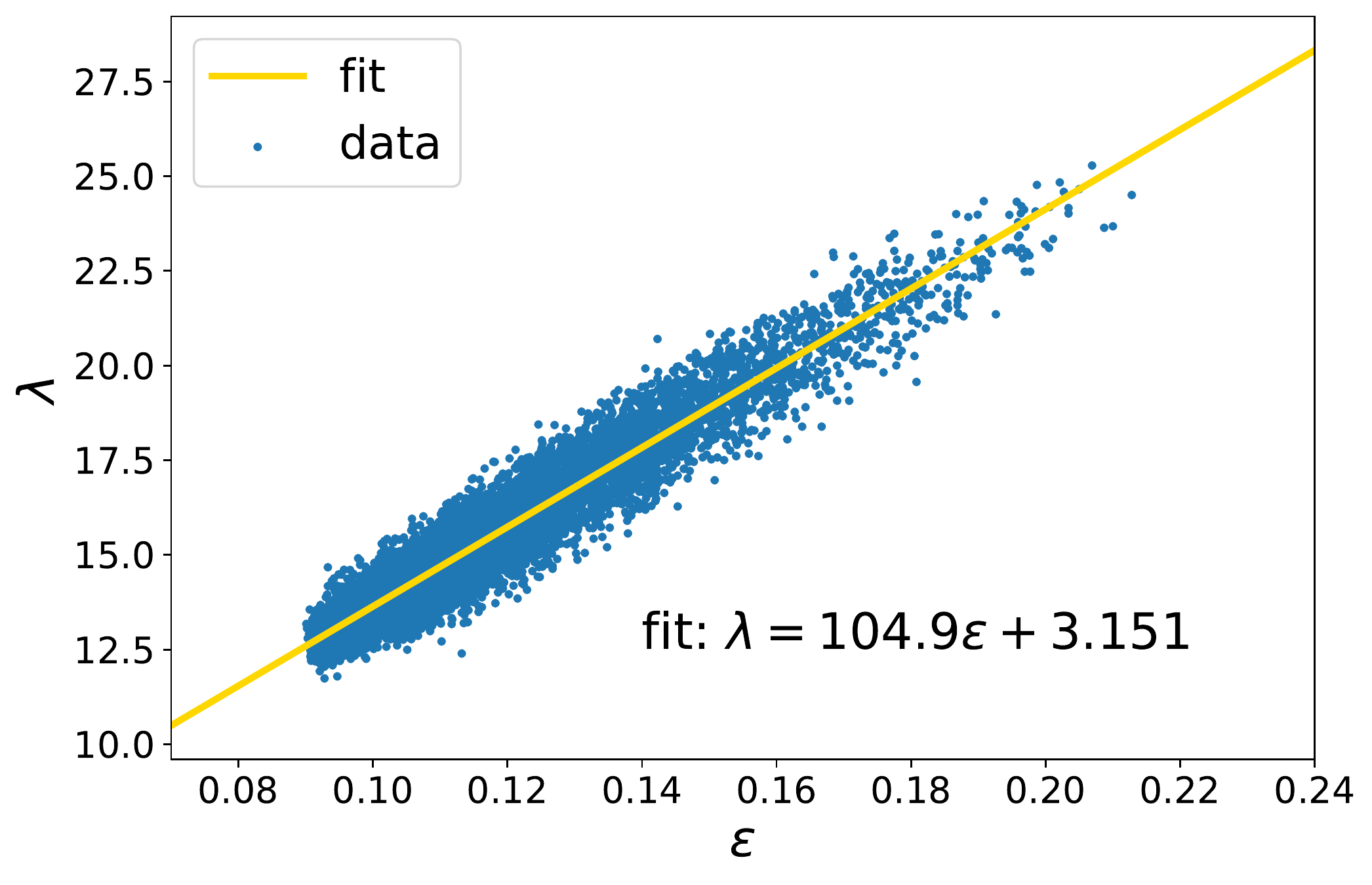}
 \caption{Same as Fig. \ref{BCyss}, but for 6dFGSv survey.}
 \label{BCyssA2}
\end{figure}

 Fig.\ref{BCyssA2} shows the relation between $\lambda$ and $\epsilon$ for 6dFGSv surveys \citep{Springob2014}. 6dFGSv is the peculiar velocity survey from the Six-degree-Field Galaxy Survey (6dFGS, \citealt{Jones2009,Jones2004}). The survey is only in southern sky, with Galactic latitude $|b|>10^{\circ}$ out to $cz \approx 16,500$ km~s$^{-1}$. The log-distance ratio of  6dFGSv sample are measured using the fundamental plane \citep{Magoulas2012}. Using Eq.\ref{lbdep}, the fit parameters are:
\be  
k=104.9~,~~b=3.151~.
\ee
This indicates that the linear relation Eq.\ref{lbdep} also exists for a Fundamental Plane survey-6dFGSv.

\section{The effects of $\delta$}\label{AP3}

\begin{figure} 
 \centering
 \includegraphics[width=100mm]{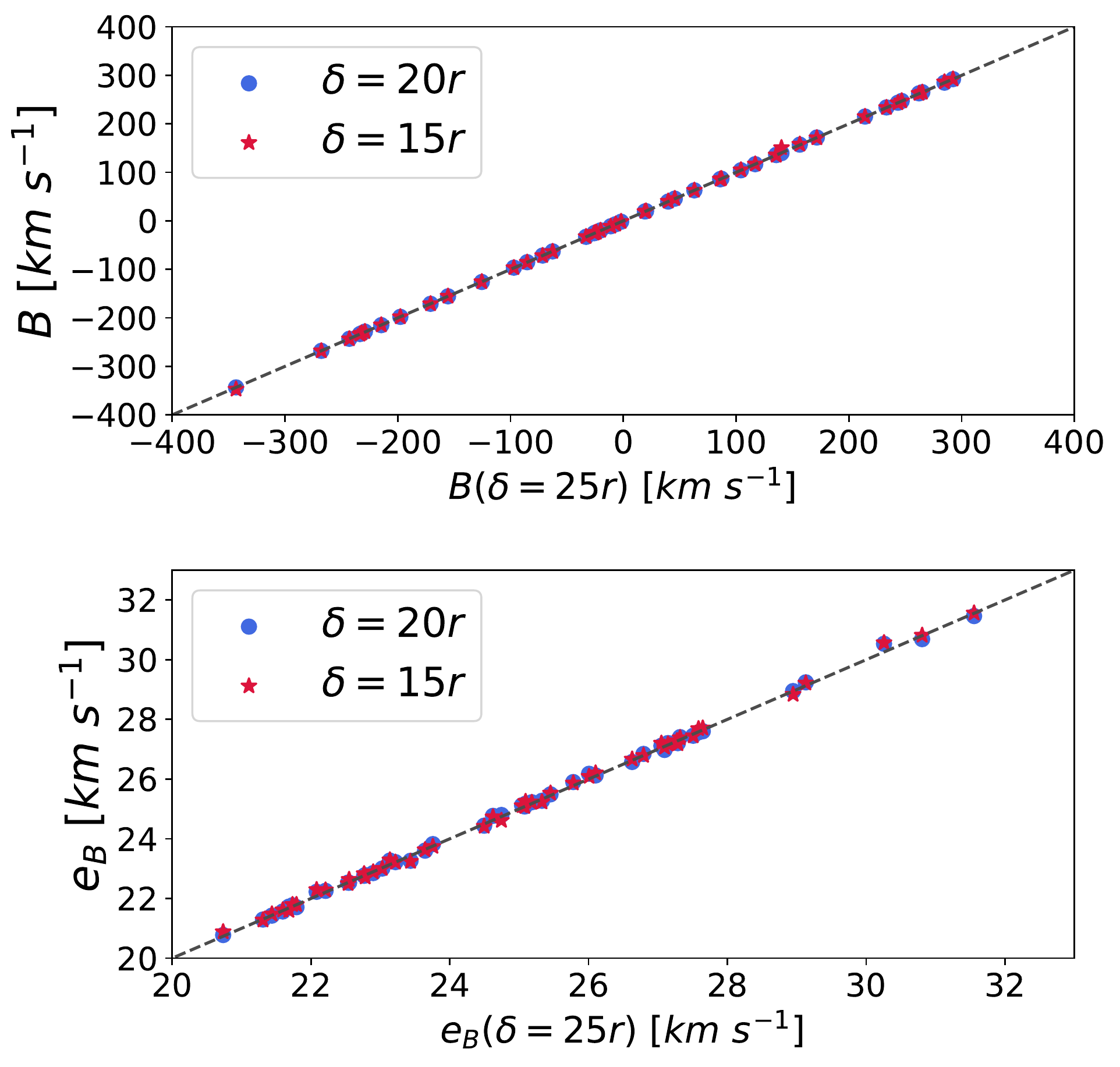}
 \caption{Top panel shows the bulk flow measurements for 16 2MTF mocks with $\delta=20r$ (blue dots) and $\delta=15r$ (red stars) against the  measurements with $\delta=25r$. The bottom panel is for the measurement errors compression.}
 \label{BCyssb2}
\end{figure}

We  choose $\delta=20r$ and $\delta=15r$ to estimate the bulk flow  for the  2MTF mocks.
The top panel of Fig.\ref{BCyssb2} shows the bulk flow measurements for 16 2MTF mocks with $\delta=20r$ (blue dots) and $\delta=15r$ (red stars) against the  measurements with $\delta=25r$. The black dashed line is the identity line. Choosing different values of $\delta$ will not change the bulk flow measurements. The bottom panel is for the measurement errors compression. Choosing different values of $\delta$ will not change the measurement errors too.

\begin{figure} 
  \centering
 \includegraphics[width=100mm]{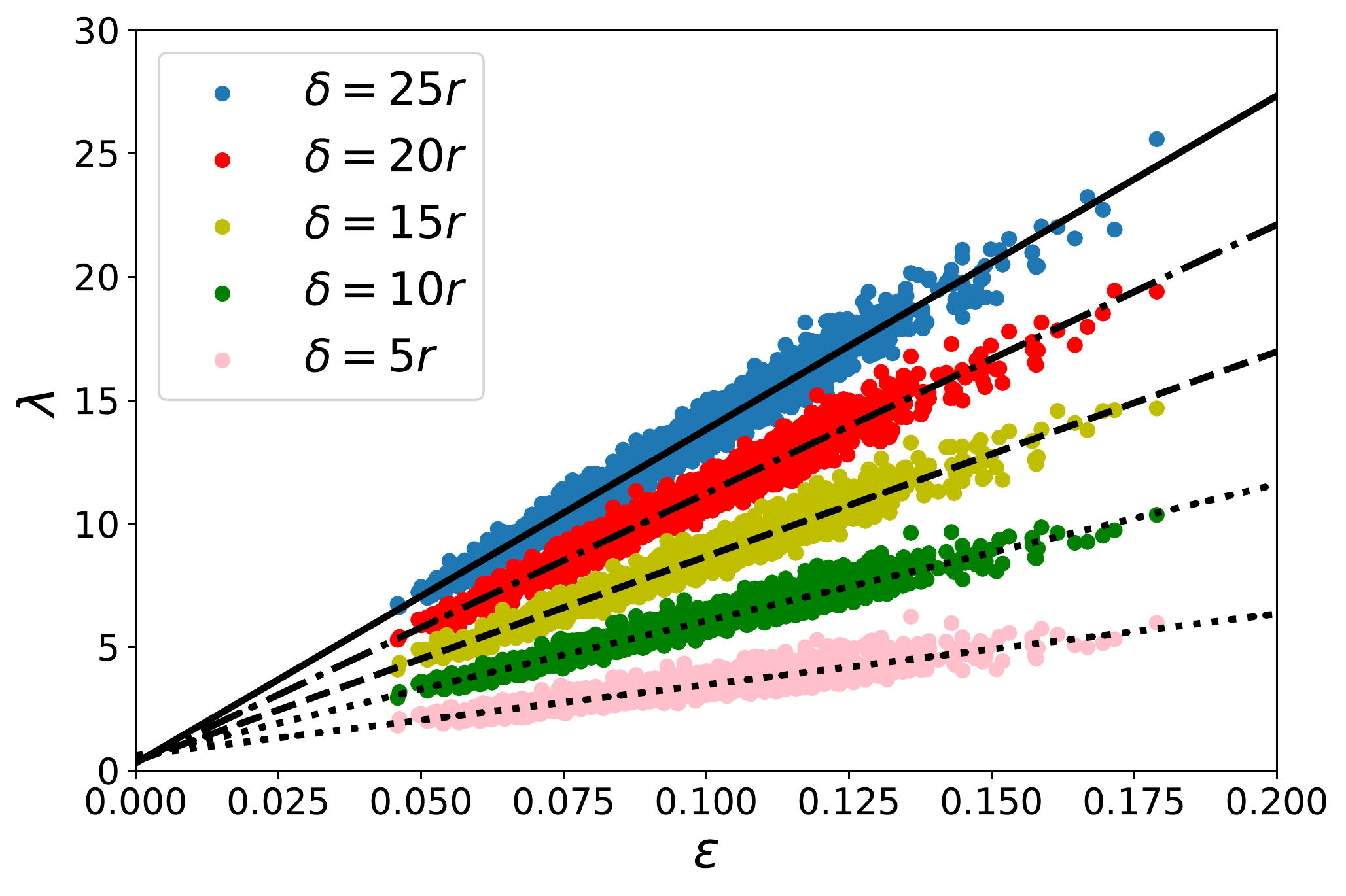}
 \caption{Same as Fig. \ref{BCyss}, but for different $\delta$. The blue, red, yellow, green and pink points are for
$\delta=25r$, $\delta=20r$, $\delta=15r$, $\delta=10r$ and $\delta=5r$, respectively. The black  lines are the best fit to the  blue, red, yellow, green and pink  points,  respectively.}
 \label{BCyssb1}
\end{figure}

We also choose different values of $\delta$ to estimate the parameter $\lambda$ for the real 2MTF survey. 
As shown in Fig.\ref{BCyssb1}, the blue, red, yellow, green and pink points are for
$\delta=25r$, $\delta=20r$, $\delta=15r$ , $\delta=10r$ and $\delta=5r$, respectively. 
The black  lines are the best fits to the  blue, red, yellow, green and pink points,  respectively. As $\delta$ increasing, the slope is increasing, while the intercept will not change too much, the fit value of $(k,b)$ is presented in Table \ref{bkvweest}. 

\begin{table}   \centering
\caption{Comparing $(k,~b)$ of different values of $\delta$.}
\begin{tabular}{|c|c|c|}
\hline
\hline
$\delta$ &$k$ &$b$\\
  &km s$^{-1}$&km s$^{-1}$\\
\hline

$25r$     & 135.1 & 0.3247 \\
\hline

$20r$     & 108.8 & 0.366 \\
\hline

$15r$     & 82.93 & 0.3939 \\
\hline

$10r$     & 55.33 & 0.5435 \\
\hline

$5r$     & 28.70 & 0.6186  \\
\hline
\end{tabular}
 \label{bkvweest}
\end{table}

 
\clearpage
\small
\bibliographystyle{mnras}
\bibliography{RAA-2021-0055-bib}

\label{lastpage}

\end{document}